\documentclass[12pt]{article}

\usepackage{amsmath,graphicx}
\usepackage{epstopdf}
\usepackage{multirow}

\usepackage{amsfonts}
\usepackage{amssymb}
\usepackage{amscd}
\usepackage{hyperref}
\usepackage[numbers, square, comma, sort&compress]{natbib}

\def\hybrid{\topmargin -20pt    \oddsidemargin 0pt
        \headheight 0pt \headsep 0pt
        \textwidth 6.25in       
        \textheight 9.5in       
        \marginparwidth .875in
        \parskip 5pt plus 1pt   \jot = 1.5ex}

\hybrid

\numberwithin{equation}{section}
\numberwithin{table}{section}

\newcommand{\be}{\begin{equation}}
\newcommand{\ee}{\end{equation}}

\newcommand{\bea}{\begin{eqnarray}}
\newcommand{\eea}{\end{eqnarray}}

\def\Fl{{\dF}}

\def\V{V}
\def\A{\Gamma}
\def\dA{a}
\def\dF{f}
\def\flux{X}
\def\U{U}

\begin{document}
\begin{titlepage}
\begin{center}
\rightline{\small ZMP-HH/11-23}
\vskip 1cm

{\Large \bf 6D Effective Action of Heterotic Compactification 

on K3 with nontrivial Gauge Bundles}	
\vskip 1.4cm
{\bf Jan Louis$^{a,b}$, Martin Schasny$^{a}$ and Roberto Valandro$^{a}$}

\vskip 0.8cm

$^{a}${\em II. Institut f\"ur Theoretische Physik der Universit\"at Hamburg, Luruper Chaussee 149, 22761 Hamburg, Germany}
\vskip 0.4cm

{}$^{b}${\em Zentrum f\"ur Mathematische Physik,
Universit\"at Hamburg,\\
Bundesstrasse 55, D-20146 Hamburg}
\vskip 0.8cm

{\tt jan.louis,martin.schasny,roberto.valandro@desy.de}

\end{center}

\vskip 1cm

\begin{center} {\bf ABSTRACT } \end{center}

\noindent
We compute the six-dimensional effective action of the heterotic
string compactified on $K3$ for  the standard embedding and for 
a class of backgrounds with line bundles and appropriate Yang-Mills
fluxes. 
We compute the couplings of the charged scalars and the bundle moduli
as functions of the geometrical $K3$ moduli from a Kaluza-Klein
analysis.
We derive the $D$-term potential and show that in the flux backgrounds 
$U(1)$ vector multiplets become massive by a St\"uckelberg mechanism. 


\vfill

December 2011

\end{titlepage}

\section{Introduction}
Heterotic model building is one 
of the possibilities to connect string theory with particle phenomenology.
The requirement of a light chiral spectrum in four space-time dimensions (4D)
together with stability arguments suggests to consider string backgrounds with
\linebreak
$\, {\mathcal N}=1$~supersymmetry in  4D. This in turn singles out  
Calabi-Yau threefolds \cite{Candelas:1985en}, appropriate $\mathbb{Z}_n$ orbifolds \cite{Dixon:1985jw}
or more generally  two-dimensional $(0,2)$ superconformal  
field theories \cite{Green:1987mn, Polchinski:1998rr} as backgrounds.

The revival of Grand Unified Theories (GUTs) in recent years resulted 
in renewed attempts to embed these field theories also in the heterotic string.
In particular field theoretic models where a GUT-group is only unbroken
in a higher-dimensional space-time background seem attractive due the 
simplicity of the Higgs-sector \cite{Kawamura:2000ev, Hall:2001pg, Hebecker:2001wq, Asaka:2001eh}.
This led to the study of  anisotropic orbifold compactification 
with an intermediate 5D or 6D effective theory \cite{Kobayashi:2004ud, Forste:2004ie, Buchmuller:2005jr, Buchmuller:2007qf}.

One of the problems of orbifold compactifications 
is the vast number of massless moduli fields. 
However, it is well known that some of them gain mass
when one considers the theory away from the orbifold point, 
i.e.\ in blown-up orbifolds or more generally in smooth Calabi-Yau backgrounds.
The relation between orbifold and smooth
Calabi-Yau compactifications 
is addressed in \cite{Honecker:2006qz,Nibbelink:2007rd,Nibbelink:2007pn,Nibbelink:2008tv,Nibbelink:2009sp,Blaszczyk:2010db,Blaszczyk:2011ig,BLSV}.
In this paper we focus instead 
on the 6D intermediate theory and derive the effective
action for smooth $K3$ compactifications from a Kaluza-Klein reduction. 
The resulting 6D effective theory has the minimal
amount of eight supercharges corresponding to  ${\mathcal N} =2$
in 4D. The scalar fields appear in tensor- and hypermultiplets 
but not in vector multiplets. 
In perturbative heterotic compactifications there is exactly one 
tensor multiplet containing the dilaton 
while all other scalars are members of hypermultiplets. 
In this case supersymmetry constrains the action to depend on a 
gauge coupling function given by the dilaton, a
quaternionic-K\"ahler metric of the hypermultiplet scalars
and a $D$-term potential \cite{Bagger:1983tt,Nishino:1986dc,Riccioni:2001bg}.

A consistent heterotic string background has to satisfy the Bianchi identity
which in turn requires a nontrivial gauge bundle on $K3$. 
As a consequence the resulting light scalar Kaluza-Klein (KK) spectrum consists of 
the moduli of $K3$, the moduli of the gauge bundle
and a set of matter fields charged under the unbroken gauge group.
For these fields we systematically compute their couplings in the effective action, extending the analysis in \cite{Green:1984bx, Walton:1987bu, Witten:1995gx, Schwarz:1995zw,
  Seiberg:1996vs, Louis:2001uy, Honecker:2006dt}.\footnote{For
  compactifications on a Calabi-Yau threefold a similar analysis has
  been performed, for example, in \cite{Dixon:1989fj, Gurrieri:2007jg, Andrey:2011np}.}
However, since the effective action sensitively depends on the choice of
the gauge bundle 
we cannot give a model-independent answer.
Instead we focus on two prominent subclasses of gauge bundles embedded in $E_8 \times E_8$: 
we discuss the well
known standard embedding of the gauge bundle into the tangent bundle
in section~\ref{SEE} and backgrounds with $U(1)$ line bundles
in section~\ref{LB}.

In the derivation of the 6D effective action we
focus on the bundle moduli and the 
matter fields and compute their  couplings as a function of the $K3$ moduli.
While low energy supersymmetry restricts the compactification manifold
to be Calabi-Yau, it also restricts the gauge bundle to be a solution of
the hermitean Yang-Mills equations (HYM)~\cite{Candelas:1985en}.
These solutions are generally constructed from a stability condition
using algebraic geometry \cite{Witten:1985bz,Friedman:1997yq,Donagi:1997dp,Donagi:1999gc,Andreas:1999ty,Andreas:2004ja}.
However, on $K3$ the HYM equations take a simple form, stating that
the background field strength is
anti-selfdual (ASD) \cite{Atiyah:1978wi, Friedman:1998tb}. 
Its massless deformations determine the light 6D particle spectrum
and lead to  ASD-preserving bundle moduli which deform the
holomorphic bundle structure and  charged matter fields
which deform the structure group embedding.

This paper is organized as follows. In section~\ref{Prelim} we set the stage 
for the later analysis and
briefly recall the multiplets and effective action
of 6D minimal supergravity (in section~\ref{6DN=1}),
and some basic facts about $K3$ (in section~\ref{K3}).
In section \ref{SEE} we then turn to the standard embedding 
and derive the effective action.
We determine the couplings of the matter fields and the bundle moduli
as a function of the $K3$ moduli.
Unfortunately for the bundle moduli these couplings can only be given in terms of
moduli-dependent
integrals on $K3$ but they are not explicitly evaluated.
As a consequence we cannot show in general that the final metric is 
quaternionic-K\"ahler as required by supersymmetry \cite{Bagger:1983tt}.
However, in an appropriate orbifold limit we show that the couplings of the matter fields in the untwisted sector are quaternionic-K\"ahler and agree with the results of \cite{Ferrara:1986qn}.
We further compute the scalar potential and show that it consistently descends from a $D$-term.



In section \ref{LB} we consider backgrounds with line
bundles \cite{Green:1984bx,Andreas:2004ja,Aldazabal:1996du,Blumenhagen:2005ga,Blumenhagen:2006ux,Honecker:2006dt, Honecker:2006qz, Kumar:2009ae,Kumar:2009zc}.
In this case the Bianchi identity is satisfied by Abelian Yang-Mills
fluxes on internal $K3$ two-cycles.
The fluxes are characterized by their group theoretical embedding
inside the Cartan subalgebra of $E_8 \times E_8$ 
and the localization inside the second cohomology lattice of $K3$.
Using a vanishing theorem we show that the resulting effective action
is consistent with 6D supergravity in that the scalar potential 
descends from a $D$-term.
We determine the couplings of the matter fields in terms of $K3$-moduli
dependent integrals.
The Abelian factors of the gauge bundle are also part of the unbroken
gauge group and the fluxes affect the effective action in two ways.
First of all, the scalars descending from the heterotic $B$-field 
get affinely gauged under the Abelian factors.
Due to the St\"uckelberg mechanism this is equivalent to the Abelian gauge
bosons becoming massive.
Second of all, in the scalar potential the (selfdual components of the) fluxes
appear as Fayet-Iliopoulos $D$-terms, leading to a stabilization of
s subset of the $K3$ moduli.
Together, for every independent gauge flux a vector multiplet and a
hypermultiplet gain a non-zero mass, consistent with the 6D anomaly
constraint.  

In appendix~\ref{def} we describe in detail the local deformation theory of gauge
connections, which is essential for the Kaluza-Klein reduction in the
main text.
In particular we establish the connection of massless internal
deformations and Dolbeault cohomology which, to our knowledge, 
is not discussed in detail in the literature.
Finally, appendix~\ref{limit} provides further details about the 
metric in the untwisted sector of the previously considered orbifold limit.


\section{Preliminaries}\label{Prelim}

In this paper we consider Kaluza-Klein reductions of the heterotic string in space-time backgrounds of the form
\begin{equation}\label{K3back}
M^{1,5}\times K3 \ ,
\end{equation}
where $M^{1,5}$ is the six-dimensional Minkowski space-time with Lorentzian signature and $K3$ is the unique compact four-dimensional Calabi-Yau manifold.
 
The starting point of the analysis is the ten-dimensional heterotic supergravity characterized by the bosonic Lagrangian\footnote{Throughout this paper we use the space-time metric signature $(-,+,+,+,...)$ and antihermitean generators for the gauge group.}
\begin{equation}\label{hetaction10}
 \mathcal{L} = \tfrac{1}{2} e^{-2\Phi}\left(R *1 + 4d\Phi \wedge * d\Phi - \tfrac{1}{3} H\wedge * H + \alpha'
 (\mbox{tr} \ F\wedge * F - \mbox{tr} \ \tilde{R} \wedge * \tilde{R})\right)\ .
\end{equation}
$\Phi$ is the ten-dimensional dilaton, $F$ is the Yang-Mills field strength in the adjoint representation of $E_8\times E_8$ and $H$ is the field strength of the Kalb-Ramond field $B$ defined as
\be\label{H}
H=dB + \alpha' (\omega^L - \omega^{YM})\ ,
\ee
where $\omega^L, \omega^{YM}$ are the gravitational and Yang-Mills Chern-Simons 3-forms, respectively. As a consequence $H$ satisfies the Bianchi identity
\begin{equation}\label{BI}
 dH = \alpha' (\mbox{tr} \ R \wedge R - \mbox{tr} \ F \wedge F)\ ,
\end{equation}
where $R$ is the Riemann curvature 2-form.\footnote{The trace
  $\mbox{tr} R \wedge R$ is evaluated in the vector representation
  $\bf 10$ of $SO(1,9)$ and $\mbox{tr} F \wedge F := \tfrac{1}{30}
  \mbox{Tr}F \wedge F$ is $\tfrac{1}{30}$ of the trace in the adjoint
  representation of $E_8 \times E_8$.} 
Finally, the last term in \eqref{hetaction10} is the Gauss-Bonnet form \cite{Zwiebach:1985uq}
\begin{equation}\label{GB}
 \mbox{tr} \tilde{R} \wedge * \tilde{R} := R_{MNPQ} R^{MNPQ} - 4
 R_{MN} R^{MN} + R^2\ .
\end{equation}

The Bianchi identity \eqref{BI} requires a nontrivial gauge bundle
over $K3$. As a consequence the original $E_8 \times E_8$  gauge group breaks
to $G$ according to 
\begin{equation}
 E_8 \times E_8 \longrightarrow G \times \langle H \rangle \ .
\end{equation}
Here $\langle H \rangle$ is the structure group of the nontrivial bundle and $G$ is the unbroken maximal commutant.

Before compactification, i.e.\ in flat ten-dimensional Minkowski
space-time $M^{1,9}$, the theory has 16 supercharges corresponding to an
${\mathcal N}=1$ supergravity in $D=10$. In a background of the form
\eqref{K3back} half of the supersymmetries are broken due to the
properties of $K3$.
Unbroken supersymmetry also constrains the gauge bundle to satisfy
the hermitean Yang-Mills equations 
\cite{Candelas:1985en}
\begin{equation}\label{hym1}
 {\mathcal F} \in H^{1,1}(K3, \mathfrak{h}), \qquad {\mathcal F} \wedge
 J = 0\ ,
\end{equation}
where $H^{1,1}(K3, \mathfrak{h})$ denotes the $(1,1)$ Dolbeault cohomology
group
with values in the adjoint bundle $\mathfrak{h}$
of $H$ and $J$ is the K\"ahler-form of
$K3$.\footnote{
Note that a solution of \eqref{hym1} also solves 
the full Yang-Mills equations.
} 
On $K3$ the hermitean Yang-Mills equations are equivalent to the
anti-selfduality condition \cite{Atiyah:1978wi, Friedman:1998tb}
\begin{equation}
 {\mathcal F} \in \Lambda^2 _-(K3, \mathfrak{h}) \ ,
\end{equation}
where $\Lambda^2 _-(K3, \mathfrak{h})$ denotes
the $-1$ eigenspace of the Hodge-$\star$ operator acting on 2-forms.
The resulting low energy effective theory is an ${\mathcal N}=1$
supergravity in $D=6$, which we shall briefly review.

\subsection{ ${\mathcal N}=1$  Supergravity in $D=6$}\label{6DN=1}

The supercharges of the $6D, {\mathcal N}=1$ supergravity form a doublet of two Weyl spinors with the same chirality, satisfying a symplectic Majorana condition. 
They are rotated into each other under the $R$ symmetry group $Sp(1)_R \cong SU(2)_R$.  The massless supermultiplets are~\cite{Townsend:1983xt}
\begin{equation}\label{multip}
\begin{aligned}
 \mbox{gravity multiplet}: \ \ \ \ &\{ g_{\mu \nu}, \psi_{\mu} ^- ,
 B_{\mu \nu} ^+ \}\ , \\
 \mbox{tensor multiplet}: \ \ \ \ &\{B_{\mu \nu} ^- , \lambda^+ , \phi
 \} \ ,\\
 \mbox{vector multiplet}: \ \ \ \ &\{\V_{\mu}, \lambda^- \}\ , \\
 \mbox{hypermultiplet}:  \ \ \ \ & \{\chi^+ , 4q \}\ ,
\end{aligned}
\end{equation}
where $ g_{\mu \nu}$ is the graviton of the six-dimensional space-time, $\psi_{\mu} ^-$ the negative chirality gravitino and $B_{\mu \nu} ^+$ is an antisymmetric tensor with 
selfdual field strength. The tensor multiplet contains a tensor $B_{\mu \nu}^-$ with anti-selfdual field strength, the dilatino $\lambda^-$ and the 6D dilaton $\phi$. The vector 
multiplet contains a gauge boson $\V_{\mu}$ and the gaugino $\lambda^+$. Finally the hypermultiplet features the hyperino $\chi^+$ together with four real scalars $q$. Note that all 
scalars, except the dilaton, are in hypermultiplets. The massless spectrum is intrinsically chiral, since the fermions of each supermultiplet have definite chirality.

The doublet structure of the 6D supercharges has further consequences for possible gauge representations. Especially, the four scalars in a hypermultiplet form a complex doublet of 
the $R$-symmetry group.\footnote{A half-hypermultiplet, which is the smallest CPT self-conjugate multiplet, can only exist, if it is in a pseudoreal gauge representation. If it is a 
gauge singlet, the two real scalars are both their own CPT-conjugate
but cannot build a $SU(2)_R$-doublet \cite{Polchinski:1998rr}.} A hypermultiplet in a complex representation $\bf R$ cannot be CPT-selfconjugate,
so hypermultiplets always occur in vector-like representations ${\bf
  R} \oplus {\bf \overline{R}}$ in the spectrum. The four scalars correspondingly group into two complex scalars in ${\bf R}$ and ${\bf \overline{R}}$, respectively.

The absence of local anomalies does not constrain the gauge group as in 10D, but rather the massless spectrum to obey \cite{RandjbarDaemi:1985wc, Erler:1993zy}
\begin{equation}\label{anomaly}
 29n_T + n_H - n_V = 273\ ,
\end{equation}
where $n_T$ denotes the number of tensor multiplets, $n_H$ the number of hypermultiplets and $n_V$ the number of vector multiplets. 
This condition is automatically satisfied in any $K3$ compactifications with supersymmetric bundle \eqref{hym1}. In this paper we only consider perturbative $K3$-compactifications where $n_T = 1$, such that $n_H - n_V = 244$ holds.

For gauge groups of the form 
\be
G = \prod_\alpha G_\alpha \times \prod_m  U(1)_m\ ,
\ee
where $G_\alpha$ denotes any simple factor and $U(1)_m$ any Abelian factor, the bosonic Lagrangian is given by
\cite{Nishino:1986dc, Riccioni:2001bg}
\begin{equation}\label{lag1}
\begin{aligned}
{\mathcal L}_6 = &\ \tfrac14 R*1 - \tfrac{1}{2} e^{-2\phi}H \wedge * H + \tfrac{1}{4} d\phi \wedge * d\phi \\
&+\tfrac{1}{2}(c_\alpha  e^{-\phi} + \tilde{c}_\alpha  e^{\phi})\mbox{tr} F^{\mathfrak{g}_\alpha} \wedge * F^{\mathfrak{g}_\alpha} - \tilde{c}_\alpha  B \wedge \mbox{tr} F^{\mathfrak{g}_\alpha} \wedge F^{\mathfrak{g}_\alpha} \\ 
&+\tfrac{1}{2}(c_{mn} \ e^{-\phi} + \tilde{c}_{mn} \ e^{\phi}) F^m \wedge * F^n - \tilde{c}_{mn} \ B \wedge F^m \wedge F^n \\
&-\tfrac{1}{2} g_{uv}(q) {\mathcal D}q^{u} \wedge * {\mathcal D}q^{v} -
V*1\ ,
\end{aligned}
\end{equation}
where the non-Abelian Yang-Mills field strengths are labeled as
$F^{\mathfrak{g}_\alpha}$ and the Abelian field strengths as
$F^m$. Due to supersymmetry, the gauge kinetic functions only depend
on the 6D dilaton $\phi$, with numerical factors $c_\alpha,
\tilde{c}_\alpha, c_{mn}, \tilde{c}_{mn}$.\footnote{It was shown recently that these numerical factors are constrained to take values in a selfdual lattice~\cite{Seiberg:2011dr}.} 
For the Abelian factors kinetic mixing, parametrized by the off-diagonal part of
$c_{mn},\tilde{c}_{mn}$ 
is possible \cite{Riccioni:1999xq}. $B$ is the sum of $B^+$ and $B^-$, and it is coupled 
to the vector multiplets via Chern-Simons forms appearing in its field strength
$H = dB + \omega^L - c_\alpha \omega^{YM} _{\mathfrak{g}_\alpha} -
c_{mn} \omega^{YM}_{mn}$, where  $\omega^L$ and
$\omega^{YM}_{\mathfrak{g}_\alpha} $ 
are standard  Chern-Simons forms while 
the ``mixed'' Abelian Chern-Simons form is given by 
\be
\label{mixed}
\omega^{YM}_{mn} = d\V^m \wedge \V^n\ .
\ee

The real hypermultiplet scalars $q^{u}, u=1,\ldots,4n_H$ constitute a
quaternionic K\"ahler target manifold ${\mathcal M}$ with metric
$g_{uv}(q)$ which only depends on the hyperscalars \cite{Bagger:1983tt}. 
The gauge group can be any isometry group of $\mathcal M$, with Killing vectors $K^{u a}$ appearing in the gauge covariant derivatives: 
\begin{equation}\label{cov1}
 {\mathcal D}q^{u} = dq^{u} - \V^a K^{u a}(q)\ ,
\end{equation}
where $a$ denotes the adjoint index of the gauge group. 

Finally, there only exists a $D$-term potential given by
\begin{equation}\label{V6}
V= - \tfrac14 \sum \limits_a \frac{(D^a)_{\ B} ^A (D^a)_{\ A} ^B}{c_\alpha
     e^{-\phi} 
+ \tilde{c}_\alpha e^{\phi}} - \tfrac14 \sum \limits_{m,n} \frac{(D^m)_{\ B} ^A (D^n)_{\ A} ^B}{c_{mn} e^{-\phi} + \tilde{c}_{mn}  e^{\phi}} \ ,
\end{equation}
where 
\begin{equation}\label{Dt}
 (D^{a,m})^A _{\ B} = \A_{u B} ^A K^{u a,m}\ , \qquad
 A,B = 1,2\ ,
\end{equation}
with ${\A}_{u B} ^A$ being a composite 
$\mathfrak{su}(2)_R$-valued connection on ${\mathcal M}$ \cite{Nishino:1986dc, Riccioni:2001bg}. 
Our main interest in the following will be to derive the 6D couplings, i.e.\ the hyperscalar metric $g_{uv}(q)$ and the explicit form of the $D$-term.

\subsection{$K3$ Compactification}\label{K3}
Before we proceed let us collect a few facts about the (unique) Calabi-Yau two-fold $K3$ (for a review see \cite{Aspinwall:1996mn}). It has a reduced holonomy
group $SU(2)_{hol}$, so its frame bundle splits as 
$SO(4) \rightarrow SU(2)_R \times SU(2)_{hol}$ into an
$SU(2)_R$ bundle which is flat over $K3$   
and the nontrivial $SU(2)_{hol}$ bundle. 
A covariantly constant spinor on $K3$ transforms as a doublet under $SU(2)_R$, so this generates the $R$-symmetry in 6D. 
Moreover, the $K3$ surface is hyper-K\"ahler and its curvature 2-form is anti-selfdual \cite{Atiyah:1978wi}. Its Hodge numbers are
\begin{equation}\label{dia}
 \begin{array}{c}
 h^{0,0}   \\  h^{1,0} \qquad h^{0,1} \\ h^{2,0} \qquad h^{1,1} \qquad h^{0,2} \\ h^{2,1}\qquad  h^{1,2} \\ h^{2,2}
\end{array} =
\begin{array}{c}
  1 \\ 0 \qquad  0 \\ 1 \qquad 20 \qquad 1 \\ 0 \qquad 0 \\ 1 
\end{array}\ .
\end{equation}
The nontrivial part is the second cohomology group $H^2(K3, \mathbb{R})$. It is a vector space of signature (3,19) with respect to the scalar product
\begin{equation}
 \langle v,w \rangle = \int v \wedge w\ , \qquad v,w\in
 H^2(K3,\mathbb{R})\ .
\end{equation}
In a basis of 2-forms $\eta_I \in H^2(K3,\mathbb{R})$ the scalar product is given by the matrix \footnote{For integral 2-forms this is the intersection matrix of the Poincar\'e dual 2-cycles \cite{Aspinwall:1996mn}.}
\begin{equation}\label{insect}
\begin{aligned}
  &\rho_{IJ} = \int \eta_I \wedge \eta_J\ ,\qquad I,J=1,\ldots,22\ .
\end{aligned}
\end{equation}
A Riemannian metric on $K3$ is defined by a positive definite three-dimensional subspace $\Sigma := H^2 _+(K3, \mathbb{R}) \subset H^2(K3, \mathbb{R})$ 
and the overall volume $\mathcal V$. Then we have the orthogonal splitting $H^2(K3) = H^2 _+(K3) \oplus H^2 _-(K3)$ and the two subspaces are eigenspaces of the Hodge $\star$-operator. 
The corresponding elements are called selfdual and anti-selfdual, respectively.

Locally the moduli space of Ricci-flat metrics takes the form \cite{Todorov}
\begin{equation}\label{K3mod}
 {\mathcal M}_{K3} = \frac{O(3,19)}{O(3) \times O(19)} \times {\mathbb R}^+\ ,
\end{equation}
which has dimension 58. A complex structure is defined by the choice of an orthonormal dreibein $\{J_s\}_{s=1,2,3} \in H^2 _+(K3, \mathbb{R})$ such that 
\begin{equation}
 J= \sqrt{2 \mathcal V} J_3 \ , \quad \Omega = J_1 + iJ_2 
\end{equation}
are the K\"ahler form and the holomorphic 2-form, respectively. They are normalized as
\begin{equation}\label{norm}
 \int J \wedge J = 2{\mathcal V} \ , \qquad \int \Omega \wedge \overline{\Omega} = 2 \ , 
\qquad \lVert \Omega \rVert^2 = \tfrac12 \Omega_{\alpha \beta} \overline{\Omega}^{\alpha \beta} = \tfrac{2}{\mathcal V} \ .
\end{equation}
The metric moduli combine with the 22 scalars $b^I$, arising from zero modes of the
Kalb-Ramond field on $K3$ to form 20 hypermultiplets in 6D.
Including the $b^I$ the geometrical moduli space given in
\eqref{K3mod} locally turns into the quaternionic-K\"ahler manifold
\cite{Seiberg:1988pf}
\begin{equation}\label{K3mod2}
 {\mathcal M} = \frac{O(4,20)}{O(4) \times O(20)}\ .
\end{equation}

\section{Standard embedding on $K3$}\label{SEE}

In the previous section we recalled that heterotic theories
have to satisfy the Bianchi identity \eqref{BI}.
For compactifications on $K3$ the integrated version yields
\begin{equation}\label{BI2}
\tfrac12 \int \limits_{K3} \mbox{tr} ({F}\wedge {F}) = \tfrac12 \int
\limits_{K3}\mbox{tr} ({R} \wedge {R}) = \chi(K3)= 24
\ ,
\end{equation}
where $\chi(K3)$ is the Euler characteristic of $K3$.
In order to preserve 6D Poincar\'e invariance all background
fields have to be tangent to $K3$. 
Then \eqref{BI2} implies that
the second Chern characters of the tangent- and Yang-Mills bundle must coincide. 
In the following we denote the Kaluza-Klein expansion around these backgrounds as
\be
A=  {\mathcal A} + \dA \ , \qquad F = {\mathcal F} + \dF \ ,\qquad
\dF = d_{\mathcal A} \dA + \tfrac12 [\dA,\dA]\ .
\ee
(We  denote background fields by calligraphic symbols such as
${\cal A}, {\cal F}, {\cal H}, {\cal R}$.) Since \eqref{BI2} is a topological equation, continuous fluctuations cannot contribute to \eqref{BI2}. 

The standard embedding is defined as the solution of \eqref{BI2} 
with the integrands identified, i.e.\ ${\mathcal F} \equiv {\mathcal
  R}$ and  
${\mathcal H} \equiv 0$ in \eqref{BI} \cite{Candelas:1985en}. 
In this case the nontrivial gauge bundle is an $SU(2)$-bundle embedded
inside one $E_8$, which is identified with the $SU(2)$
structure-bundle associated 
with the holomorphic tangent bundle ${\mathcal T}_{K3}$. 
The standard embedding breaks one $E_8$ to the maximal commutant $E_7$, i.e.
\begin{equation}\label{SE}
 E_8 \times E_8 \longrightarrow E_8 \times E_7 \times \langle SU(2) \rangle \ ,
\end{equation}
where $\langle H \rangle$ denotes the broken group factor. For the standard embedding the hermitean Yang-Mills equations
\eqref{hym1} take the form 
\begin{equation}\label{hym2}
 {\mathcal F} \in H^{1,1}(\mbox{End} \ {\mathcal T}_{K3}) \ , \ \ \ \
 {\mathcal F} \wedge J = 0 \ ,
\end{equation}
where $\mbox{End} \ {\mathcal T}_{K3}$ is the bundle of linear
transition functions on ${\mathcal T}_{K3}$, 
i.e.\ locally $\mathfrak{su}(2)$ valued matrix functions.
Note that ${\mathcal F}$ is automatically anti-selfdual
since the $K3$-curvature is.

\subsection{Reduction of the Yang-Mills sector}
All bosonic charged matter multiplets arise from zero modes of the 10D vector
fields~$A$ of the broken $E_8$. 
The massless fields are determined by deformations of the background
gauge connection $A = {\mathcal A} + \dA$.
Group theoretically $\dA$ transforms in the ${\bf 10}$-dimensional
representation 
of the Lorentz group $SO(9,1)$ and in the ${\bf 248}$-dimensional
adjoint representation of $E_8$.
Decomposing
the ${\bf 248}$ under $E_8 \to E_7 \times  SU(2)$ we have
\begin{equation}\label{dec2}
 {\bf 248} \rightarrow ({\bf 133},{\bf 1}) \oplus ({\bf 1},{\bf 3})
 \oplus({\bf 56},{\bf 2})\ ,
\end{equation}
while decomposing the ${\bf 10}$ under   $SO(9,1)\to SO(5,1) \times SO(4)$
yields
\begin{equation}
 {\bf 10} \rightarrow ({\bf 6},{\bf 1}) \oplus ({\bf 1},{\bf 4}) \ .
\end{equation}
In terms of the gauge potential we denote the latter split by 
$a = a_1 + a_{\bar 1}$ where  $a_1$ denotes a one-form on $M^{1,5}$
while $a_{\bar 1}$ is an `internal' one-form on $K3$.

The non-linearity of the free 10D Yang-Mills equation complicates 
the determination of the massless modes in the Kaluza-Klein procedure. 
In appendix A we perform the Kaluza-Klein reduction in detail
and show that generically 
the scalar zero modes 
are in the cohomology $H^{0,1}(K3, E)$, where $E$ is a 
bundle associated with the right entries in the decomposition 
\eqref{dec2}.\footnote{This result is usually derived counting 
zero modes of the Dirac operator and then using  supersymmetry.
In appendix~A we rederive this result directly from the deformation
of the gauge connection.}
The result is
\begin{equation}\label{a1}
 a_1 = \V^{\bf 133} \ , \qquad a_{\bar 1} = C^{\bf 56} _j \omega_j ^{\bf 2} + \overline{C}^{\overline{\bf 56}} _j \overline{\omega}_j ^{\overline{\bf 2}} 
+ \xi_k \alpha_k ^{\bf 3} + \overline{\xi}_k \overline{\alpha}_k ^{\bf 3}\ ,
\end{equation}
where $V^{\bf 133}$ is the 6D gauge potential of the unbroken
$E_7$. The $C^{\bf 56} _j$ are complex charged scalars and 
$\xi_k$ are complex singlet scalars, called bundle moduli. 
The latter are deformations that preserve the ASD condition of the background.
Their multiplicities are
given by the cohomology groups of their corresponding zero modes
\begin{equation}\label{mult}
\begin{aligned}
  \omega_j ^{\bf 2} = (\omega_j)_{\bar \alpha} ^{\beta} dz^{\bar \alpha} \in H^{0,1} ({\mathcal T}_{K3}),& \qquad  j= 1,...,20 \ ,\\
  \alpha_k ^{\bf 3} = (\alpha_k)^s _{\bar \alpha} dz^{\bar \alpha} \in H^{0,1} (\mbox{End} \ {\mathcal T}_{K3}),& \qquad k = 1,...,90 \ .
\end{aligned}
\end{equation}
The $\omega_j ^{\bf 2}$ and  $\alpha_k ^{\bf 3}$ are one-forms which take values
in the vector bundles  $E_{\bf 2} \cong {\mathcal T}_{K3}$ and 
$E_{\bf 3}\cong~\mathfrak{su}(2)\subset\mbox{End} \ {\mathcal T}_{K3}$, respectively. This is denoted by the indices $\beta = 1,2$ and $s=1,2,3$ in \eqref{mult}.
Note that the ${\bf 3} = \mathfrak{su}(2)$ is a real representation while $\bf 56$ and $\bf 2$ are both pseudoreal.  
Therefore the 20 complex scalars $C_j ^{\bf 56}$ align in 20
half-hypermultiplets, or equivalently 10 hypermultiplets. The 90
complex bundle moduli align in 45 hypermultiplets and 20 additional 
hypermultiplets 
arise from the 58 geometrical moduli combined with the 22 Kalb-Ramond axions. 
The second $E_8$ remains unbroken and yields a 6D pure Yang-Mills hidden sector with one vector multiplet in the $\bf 248$. The constraint for anomaly freedom \eqref{anomaly} is fulfilled 
as follows:
\begin{equation}
 n_V = 133 + 248 = 381 \ , \ \ \ n_H = 10 \cdot {56} + 45 + 20  = 625\ .
\end{equation}

From \eqref{a1} we derive the Kaluza-Klein expansion of the Yang-Mills field strength, 
\begin{equation}\label{Fse}
 \dF = \dF_2 ^{({\bf 133},{\bf 1})} + \dF_{1,\bar{1}} ^{({\bf 1},{\bf 3})} + \dF_{1,\bar{1}} ^{({\bf 56},{\bf 2})}  + \dF_{\bar 2} ^{({\bf 1}, {\bf 3})} + \dF_{\bar 2} ^{({\bf 133}, {\bf 1})} \ .
\end{equation}
Here and in the following, we write $\Fl_{R, \bar{S}}$ for an
$(R+S)$-form with $R$ external (6D space-time) and $S$ internal ($K3$) indices. The different terms are orthogonal with respect to the scalar product $\langle F, G \rangle = 
\mbox{tr} F \wedge * G$. 
The first term in \eqref{Fse} is the 6D field strength of the unbroken $E_7$ 
\begin{equation}
\Fl_2 ^{({\bf 133},{\bf 1})} = dV^{\bf 133} + \tfrac12 [V^{\bf 133}, V^{\bf 133}] \ . 
\end{equation}
The next two terms in \eqref{Fse} are given by
\begin{equation}
\begin{aligned}
\label{kin1}
 \dF_{1,\bar{1}} ^{({\bf 1},{\bf 3})} &= d\xi_k \wedge \alpha^s _k + d\overline{\xi}_k \wedge \overline{\alpha}^s _k \ , \\
\dF_{1,\bar 1} ^{({\bf 56},{\bf 2})} &=  {\mathcal D}C_j ^x \wedge \omega_j ^{\beta} + {\mathcal D}\overline{C}_j ^{\bar x} \wedge \overline{\omega}_j ^{\bar \beta} \ , 
\end{aligned}\end{equation}
where we label the ${\bf 56}$ by the index $x=1,\ldots,56$.
In this notation the $E_7$-covariant derivative reads ${\mathcal D}
C_i ^{x} = dC_i ^x + V^a (\tau^a)_y ^{\ x} C_i ^y$ with $\tau^a$ being
the $E_7$ generator. 
Finally, let us derive the last two terms $\Fl_{\bar 2}$ in \eqref{Fse}.
Using the zero-mode property $d_{\mathcal A} a_{\bar 1} = 0$ (derived in appendix A) we obtain
\begin{equation}\label{comm}
 \Fl_{\bar 2} = \Bigl[\overline{\xi}_k \overline{\alpha}_k ^{\bf 3}, \ \xi_k \alpha_k ^{\bf 3} \Bigr]
 + \tfrac12 \Bigl[C^{\bf 56} _j \omega_j ^{\bf 2} + \overline{C}^{\overline{\bf 56}} _j \overline{\omega}_j ^{\overline{\bf 2}}, \
 C^{\bf 56} _j \omega_j ^{\bf 2} + \overline{C}^{\overline{\bf 56}} _j \overline{\omega}_j ^{\overline{\bf 2}}\Bigr] \ .
\end{equation}
The first commutator transforms in the $({\bf 1}, {\bf 3})$
representation. Furthermore we show in appendix~A that it 
preserves the hermitean Yang-Mills equations \eqref{hym2} and therefore
can be viewed as  a flat deformation of the background field strength
 $\delta {\mathcal F}$.
The second commutator results in two representations
\begin{equation}\label{comm2}
\begin{aligned}
 ({\bf 56}, {\bf 2}) \otimes_A ({\bf 56}, {\bf 2}) &= ({\bf 1}_A ,
 {\bf 3}_S) \oplus ({\bf 133}_S, {\bf 1}_A) \ , 
\end{aligned}
\end{equation}
which in terms of generators amounts to
\begin{equation}\label{comm2p}
\begin{aligned}
 {} [T_{x\alpha}, T_{y\beta}]   &= \varepsilon_{xy} \sigma_{\alpha \beta} ^s T_s + \tau^a _{xy} \varepsilon_{\alpha \beta} T_a \ , \\
 [T_{x\alpha}, \overline{T}_{\bar{y} \bar{\beta}}] &= \varepsilon_{x\bar{y}} \sigma_{\alpha \bar{\beta}} ^s T_s + \tau^a _{x\bar{y}} h_{\alpha \bar{\beta}} T_a \ .
\end{aligned}
\end{equation}
$\varepsilon_{xy}$ and $\varepsilon_{\alpha \beta}$ are the invariant
antisymmetric tensors of $E_7$ and $SU(2)$ respectively. 
$\sigma^s _{\alpha \beta}$ are the Pauli matrices and $\tau^a _{xy}$ the $E_7$-generators in the $\bf 56$-representation. 
Since we  have the complex conjugated fields in \eqref{a1} we also need the second commutator with different invariant tensors: For the $\tau^a _{x \bar{y}}$ to be again antihermitean, 
$\overline{\tau^a _{x \bar{y}}} = -\tau^a _{y \bar{x}}$, the tensor $h$ must satisfy 
\begin{equation}\label{kaehler}
\overline{h_{\alpha \bar{\beta}}} = h_{\beta \bar{\alpha}} \ .
\end{equation}
However, since the commutator in \eqref{comm} is a product of global 1-forms, the result is a global 2-form on $K3$. Therefore the invariant tensors 
$\sigma^s _{\alpha \beta}$, $\varepsilon_{\alpha \beta}$ and
$h_{\alpha \beta}$ must be
extended to global tensors on $K3$. In fact, 
\eqref{kaehler} is the property of a K\"ahler metric and $\varepsilon_{\alpha \beta}$ is a local expression of the holomorphic 2-form. Hence, we set
\begin{equation}\label{ext}
\begin{aligned}
 h_{\alpha \bar{\beta}} &\longrightarrow \tfrac{1}{\sqrt{2 \mathcal V}} g_{\alpha \bar{\beta}} \ , \\
 \varepsilon_{\alpha \beta} &\longrightarrow \Omega_{\alpha \beta} \ , \\
 \sigma^s _{\alpha \beta} &\longrightarrow \sigma^s _{\alpha \beta} \in \Lambda^2 (\mbox{End} \ {\mathcal T}_{K3}) \ .
\end{aligned}
\end{equation}
Since the $\bf 56$ is pseudoreal we will omit the bar on the indices $\bar{x}, \bar{y}$ in the following. 
With this we get
\begin{equation}
 \Fl_{\bar 2} = \delta {\mathcal F} + \Fl_{\bar 2} ^{({\bf 1}, {\bf 3})} + \Fl_{\bar 2} ^{({\bf 133}, {\bf 1})} \ ,
\end{equation}
where 
\begin{equation}\label{problem}
 \Fl_{\bar 2} ^{({\bf 1},{\bf 3})} = 
\begin{pmatrix} \overline{C}^{x} _i \\ C^x _i \end{pmatrix}^T \   
\begin{pmatrix} \sigma^s _{\bar{\alpha} \beta} \overline{\omega} ^{\bar \alpha} _i \wedge \omega^{\beta} _j & \sigma^s _{\bar{\alpha} \bar{\beta}} \ 
\overline{\omega}^{\bar \alpha} _i \wedge \overline{\omega}^{\bar \beta} _j \\ \sigma^s _{\alpha \beta} \ \omega^{\alpha} _i \wedge \omega^{\beta} _j & 
\sigma^s _{\alpha \bar{\beta}} \ \omega^{\alpha} _i \wedge \overline{\omega}^{\bar \beta} _j \end{pmatrix} 
\varepsilon_{xy} \begin{pmatrix} C^y _j \\ \overline{C}^{y} _j \end{pmatrix} \ ,
\end{equation}
\begin{equation}\label{Dse}
\Fl_{\bar 2} ^{({\bf 133},{\bf 1})} = \begin{pmatrix} \overline{C}^{x} _i \\ C^x _i \end{pmatrix}^T \   \begin{pmatrix} \tfrac{1}{\sqrt{2 \mathcal V}} 
g_{\bar{\alpha} \beta} \ \overline{\omega} ^{\bar \alpha} _i \wedge \omega^{\beta} _j & \overline{\Omega}_{\bar{\alpha} \bar{\beta}} \ 
\overline{\omega}^{\bar \alpha} _i 
\wedge \overline{\omega}^{\bar \beta} _j \\ \Omega_{\alpha \beta} \ \omega^{\alpha} _i \wedge \omega^{\beta} _j & \tfrac{1}{\sqrt{2 \mathcal V}} 
g_{\alpha \bar{\beta}} \ \omega^{\alpha} _i 
\wedge \overline{\omega}^{\bar \beta} _j \end{pmatrix} (\tau^a) _{xy} \begin{pmatrix} C^y _j \\ \overline{C}^{y} _j \end{pmatrix} \ . 
\end{equation}
We included the factor $\tfrac{1}{\sqrt{2 \mathcal V}}$ in \eqref{ext} such that all matrix elements of the final expression are independent of the $K3$-volume.

\subsection{Reduction of the Kalb-Ramond sector}\label{KR1}

We now turn to the reduction of the $H \wedge * H$-term in the 10D 
Lagrangian~\eqref{hetaction10} where $H=dB + \alpha' (\omega^L - \omega^{YM})$
is a gauge invariant and thus globally defined 3-form. In the
KK-reduction $H$ splits into two pieces
\begin{equation}\label{Hintt}
 {H} \longrightarrow H_3 \ + \ H_{1,\bar{2}}  \ ,
\end{equation}
where $H_3$ is the standard  6D Kalb-Ramond term with all indices 
in the space-time direction. This term reduces straightforwardly 
yielding the second term in \eqref{lag1}. For  $H_{1,\bar{2}}$ on the
other hand we need to perform the KK-reduction
with more care.

Let us start by considering the Yang-Mills Chern-Simons form 
which in 10D is defined by                             
$\omega^{YM} = \mbox{tr} (F \wedge A) - \tfrac{1}{3}\mbox{tr} (A \wedge A \wedge A)$.
For the $\omega_{1,\bar{2}}$ component we then have 
\begin{equation}
 \omega_{1,\bar{2}} ^{YM} = \mbox{tr}(\Fl_{1,\bar 1} \wedge
 A_{\bar 1}) + \mbox{tr}(F_{\bar 2} \wedge a_1) - \mbox{tr}(A_{\bar 1}
 \wedge A_{\bar 1} \wedge a_1) \ .
\end{equation}
Inserting the Kaluza-Klein expansions \eqref{a1} and \eqref{Fse}, including the background fields $A_{\bar 1} = {\mathcal A} + a_{\bar 1}$, $F_{\bar 2} = {\mathcal F} + \delta{\mathcal F}
 + \Fl_{\bar 2}$, the nonvanishing terms are 
\begin{equation}\label{CSint}
 \omega_{1,\bar{2}} ^{YM} = \mbox{tr}\Bigl(\Fl_{1, \bar 1} ^{({\bf 56}, {\bf 2})} \wedge a_{\bar 1} ^{({\bf 56}, {\bf 2})}\Bigr) 
+ \mbox{tr} \Bigl( \Fl_{1, \bar 1} ^{({\bf 1}, {\bf 3})} \wedge ({\mathcal A} + a_{\bar 1} ^{({\bf 1}, {\bf 3})}) \Bigr) \ .
\end{equation}
Similar to the commutators \eqref{comm2}, the traces of antihermitean generators yield invariant tensors that are extended to global tensors on $K3$
\begin{equation}\label{ext1}
\begin{aligned}
-\mbox{tr}(T_s T_t)&=\delta_{st} \longrightarrow h_{st} \ , \\
 -\mbox{tr}(T_{x\alpha}T_{y\beta})&=\varepsilon_{xy}\varepsilon_{\alpha \beta} \longrightarrow \varepsilon_{xy}\Omega_{\alpha \beta} \ , \\
-\mbox{tr}(T_{x\alpha}\overline{T}_{{y}\bar{\beta}})&=\delta_{xy} h_{\alpha \bar{\beta}} \longrightarrow \tfrac{1}{\sqrt{2 \mathcal V}} 
\delta_{xy} g_{\alpha \bar{\beta}} \ .
\end{aligned} 
\end{equation}
Here $h_{st}$ is a hermitean metric on the adjoint $\mbox{End} \
{\mathcal T}_{K3}$-bundle. Inserting \eqref{ext1} and \eqref{kin1} into \eqref{CSint} we arrive at
\begin{equation}\label{chern}
\begin{aligned}
 \omega_{1,\bar{2}} ^{YM} =&- \begin{pmatrix} {\mathcal D}\overline{C}_i ^x \\ {\mathcal D} C_i ^x \end{pmatrix}^T \begin{pmatrix} \tfrac{1}{\sqrt{2 \mathcal V}} \delta_{xy} g_{\bar{\alpha} \beta} \overline{\omega}_i ^{\bar \alpha} 
\wedge \omega_j ^{\beta} & \varepsilon_{xy} \overline{\Omega}_{\bar{\alpha} \bar{\beta}} \overline{\omega}_i ^{\bar \alpha} \wedge 
\overline{\omega}_j ^{\bar \beta} \\ \varepsilon_{xy} \Omega_{\alpha \beta} \omega_i ^{\alpha} \wedge \omega_j ^{\beta} & \tfrac{1}{\sqrt{2 \mathcal V}} \delta_{xy} g_{\alpha {\bar \beta}} \omega_i ^{\alpha} \wedge 
\overline{\omega}_j ^{\bar \beta} \end{pmatrix} \begin{pmatrix} C_j ^y \\ \overline{C}_j ^y \end{pmatrix} \\
-&\begin{pmatrix} d\overline{\xi}_k \\ d\xi_k \end{pmatrix}^T \begin{pmatrix} h_{st} \overline{\alpha}_k ^s \wedge \alpha_l ^t & 
    h_{st} \overline{\alpha}_k ^s \wedge \overline{\alpha}_l ^t \\ h_{st} \alpha_k ^s \wedge \alpha_l ^t & h_{st} \alpha_k ^s \wedge \overline{\alpha}_l ^t \end{pmatrix}
    \begin{pmatrix} \xi_l \\ \overline{\xi}_l  \end{pmatrix} - d\xi_k \left(h_{st} \alpha_k ^s \wedge \mathcal{A}^t \right) - 
    d\overline{\xi}_k \left(h_{st} \overline{\alpha}_k ^s \wedge \mathcal{A}^t \right).
\end{aligned}
\end{equation}
In \eqref{N1} and \eqref{N4} we show that the zero modes of the first
two terms can be written in terms of harmonic 2-forms and
thus are globally defined.
 The last two terms on the other hand 
contain the gauge connection
 $\mathcal A$ explicitly and therefore 
are gauge-variant and  globally not well defined.
However, they are a total derivative in 6D and thus can be absorbed
into $dB_{1,\bar{2}}$ by redefining $B_{\bar{2}}$.
This has the additional benefit that after the redefinition
$B_{1,\bar{2}}$ is also gauge invariant which follows from the fact that 
$H$ and the first two
terms of  $\omega_{1,\bar{2}} ^{YM}$ in \eqref{chern} are gauge
invariant.
Therefore the internal redefined $B_{\bar 2}$-field can be expanded globally as
\begin{equation}\label{inv}
 B_{\bar 2} = b^I \eta_I 
\ .
\end{equation}
Finally let us note that  the Lorentz Chern-Simons form 
$\omega^L _{1, \bar{2}}$ also is a total space-time derivative in 6D and can
similarly be 
absorbed into a redefinition of $B_{\bar 2}$.
Thus altogether we have 
\begin{equation}\label{HH}
\begin{aligned}
 H_{1, \bar 2} =&\ dB_{\bar 2} \ + \ \alpha' \omega^{CS} _{1, \bar 2}\\ 
=& \ db^I \wedge \eta_I
- \alpha' \begin{pmatrix}  {\mathcal D}\overline{C}_i ^x \\ {\mathcal D}C_i ^x \end{pmatrix}^T \begin{pmatrix} \tfrac{1}{\sqrt{2 \mathcal V}} \delta_{xy} g_{\bar{\alpha} \beta} \overline{\omega}_i ^{\bar \alpha} 
\wedge \omega_j ^{\beta} & \varepsilon_{xy} \overline{\Omega}_{\bar{\alpha} \bar{\beta}} \overline{\omega}_i ^{\bar \alpha} \wedge 
\overline{\omega}_j ^{\bar \beta} \\ \varepsilon_{xy} \Omega_{\alpha \beta} \omega_i ^{\alpha} \wedge \omega_j ^{\beta} & \tfrac{1}{\sqrt{2 \mathcal V}} \delta_{xy} g_{\alpha {\bar \beta}} \omega_i ^{\alpha} \wedge 
\overline{\omega}_j ^{\bar \beta} \end{pmatrix} \begin{pmatrix} C_j ^y \\ \overline{C}_j ^y \end{pmatrix} \\
&\qquad -\alpha' \begin{pmatrix} d\xi_k \\ d\overline{\xi}_k \end{pmatrix}^T \begin{pmatrix} h_{st} \alpha_k ^s \wedge \overline{\alpha}_l ^t & 
    h_{st} \alpha_k ^s \wedge \alpha_l ^t \\ h_{st} \overline{\alpha}_k ^s \wedge \overline{\alpha}_l ^t & h_{st} \overline{\alpha}_k ^s \wedge \alpha_l ^t \end{pmatrix}
    \begin{pmatrix} \overline{\xi}_l \\ \xi_l \end{pmatrix} \ .
\end{aligned}
\end{equation}

\subsection{6D Effective action}\label{kin}
Using the results from the previous sections we now derive the 6D effective action, first focusing on the kinetic terms. The effective action of the gravity-dilaton sector has been determined in
ref.~\cite{Louis:2009dq} 
and we include their result in the following. In the Einstein frame the 6D dilaton $\phi$ has to be defined as 
\begin{equation}
 \phi = \Phi - \tfrac12 \mbox{ln} {\mathcal V} \ ,
\end{equation}
where $\Phi$ is the 10D dilaton and $\mathcal V$ is the $K3$
volume. The Einstein-frame metric is given by $g_{\mu \nu} = e^{-\phi} g_{\mu \nu} ^{(10)}$. 
From this redefinition one gets a factor of ${\mathcal V}^{-1}$ in
front of all terms in the Lagrangian with nontrivial $K3$ integral.
Altogether we get
\begin{equation}\label{Kinetic6Dterms}
\begin{aligned}
 {\mathcal L}_6 = & \ \tfrac12 R * 1 \ - \ \tfrac16 e^{-2\phi}  H
 \wedge * H \ + \ \tfrac{\alpha'}{2} e^{-\phi} \mbox{tr} F^{\bf 133}
 \wedge * F^{\bf 133}
+ \tfrac92 d\phi \wedge * d\phi \ \\
 & - \ \tfrac{\alpha'}{\mathcal V} {\mathcal G}_{kl} d\overline{\xi}_{k}
 \wedge* d\xi_l \ + \ \tfrac14 h_{IJ} dt_s ^I \wedge * dt_s ^J \ - \ \tfrac{1}{8{\mathcal V}^2} d{\mathcal V} \wedge * d{\mathcal V} \\
 & - \ \alpha' G_{ij} \delta_{x y} {\mathcal D}\overline{C}^{x} _{i} \wedge * 
 {\mathcal D}C^y _j \ - \ \tfrac1{6\mathcal V} g_{IJ} {\mathcal D}_c b^I \wedge * {\mathcal D}_c b^J - V * 1\ ,
\end{aligned}
\end{equation}
where the $t_s ^I$ are the  $K3$ moduli
which, together with the volume, span the moduli space \eqref{K3mod} with
the metric denoted by $h_{IJ}$.\footnote{For an explicit expression of 
$h_{IJ}$ see, for example, \cite{Louis:2009dq}. Note that from the classical 10D supergravity we cannot deduce the 6D Green-Schwarz term and the full dilaton couplings of \eqref{lag1}.}
The charged scalars are gauged under the unbroken $E_7$ via the covariant derivative 
\begin{equation}\label{covV}
 {\mathcal D}C_i ^{x} = dC_i ^x + \V^a (\tau^a)_y ^{\ x} C_i ^y \ .
\end{equation}
For the $b$-scalars we have
\begin{equation}\label{Db}
\begin{aligned}
 {\mathcal D}_c b^I = db^I &- \alpha'\delta_{xy} M^I _{ij} \overline{C}^x _i \overleftrightarrow{\mathcal D} C^y _j - \alpha'\varepsilon_{xy} (N^I _{ij} C^x _i {\mathcal D} C^y _j + c.c.) \\
&- \alpha' {\mathcal M}^I _{k l} \overline{\xi}_{k} \overleftrightarrow{d} \xi_l - \alpha' ({\mathcal N}^I _{kl} \xi_k d \xi_l + c.c.) \ .
\end{aligned}
\end{equation}
Here $\overline{\xi}_k \overleftrightarrow{d} \xi_l := \overline{\xi}_k d\xi_l
- \xi_l d \overline{\xi}_k$ is the skew-symmetric derivative and we use
the same definition for the 
$E_7$-covariant derivative $\overleftrightarrow{\mathcal D}$. 
The scalar couplings in \eqref{Kinetic6Dterms} 
depend on the $K3$ moduli and are given by
\begin{equation}\label{smetric}
\begin{aligned}
 {\mathcal G}_{kl} &=  \int h_{st} \overline{\alpha}^s _{k} \wedge
 \star \alpha^t _l \ , \qquad  
G_{ij} = \tfrac{\gamma_i \gamma_j}{2\sqrt{2 \mathcal V}}g_{ij} \ , \qquad   
  g_{IJ} = \int \eta_I \wedge \star \eta_J \ .
\end{aligned}
\end{equation}
The coupling $G_{ij}$ of the charged scalars (no summation over $i,j$ implied) is proportional to $b$-scalar coupling $g_{IJ}$, 
restricted to $H^{1,1}(K3, \mathbb{R})$. Moreover, it contains the moduli dependent functions
\begin{equation}
\gamma_i = \frac{{\mathcal V}^{\frac14}}{(\int J \wedge \eta_i)^{\frac12}} \ .
\end{equation}
We find that these are necessary in the charged zero mode isomorphy
\eqref{zero1}, in order to match with the orbifold limit known from
\cite{Ferrara:1986qn}. 
${\mathcal G}_{kl}$ is the metric on the space of ASD connections.
All couplings are derived in more detail in
the appendices \ref{zeroSE} and \ref{potSE}.

The coupling functions appearing in \eqref{Db} read
\begin{equation}\label{N}
\begin{aligned}
 N^I _{ij} &= \int \Omega_{\alpha \beta} \omega^{\alpha} _i \wedge \omega^{\beta} _j \wedge \eta^I 
 = \tfrac12 \gamma_i \gamma_j \rho_{ij} \rho^{IJ} (\langle J_1, \eta_J \rangle - i \langle J_2, \eta_J \rangle) \ ,\\
 M^I _{ij} &= \tfrac{1}{\sqrt{2 \mathcal V}} \int g_{\bar{\alpha} \beta} \overline{\omega}^{\bar \alpha} _{i} \wedge \omega^{\beta} _j \wedge \eta^I 
= \tfrac{i}{2} \gamma_i \gamma_j \rho^{IJ} \Bigl( \rho_{ij} \langle J_3, \eta_J \rangle - \langle J_3, \eta_i \rangle \rho_{jJ} - \langle J_3, \eta_j \rangle \rho_{iJ} \Bigr) \ ,
\end{aligned} 
\end{equation}
(no summation over $i,j$ implied) and are derived in  \eqref{N3} and \eqref{N1}. 
Here $\langle \cdot, \cdot \rangle$ is the scalar product on $H^2(K3, \mathbb{R})$ and $\rho_{ij}$ is
the $K3$ intersection matrix restricted to $H^{1,1}(K3,\mathbb{R})$. 
Since the definition of $H^{1,1}(K3,\mathbb{R})$ depends on a choice
of the complex structure $\rho_{ij}$ depends on the $K3$ moduli.
For the couplings of the $\xi_k$ in \eqref{Db} we find
\begin{equation}
\begin{aligned}
 {\mathcal M}^I _{kl} &= \rho^{IJ} \int h_{st} \overline{\alpha}^s _k
 \wedge {\alpha}^t_l
 \wedge \eta_J
=\rho^{IJ} \rho_{iJ} c^i _{kl} \ , \\
{\mathcal N}^I _{kl} &= \rho^{IJ} \int h_{st} \alpha_k ^s \wedge \alpha_l ^t 
 \wedge \eta_J=
e_{kl} (\langle \eta_J, J_1 \rangle - i \langle \eta_J, J_2 \rangle) \ ,  
\end{aligned}
\end{equation}
where we defined $c^i _{kl}, e_{kl}$ 
as the  (antisymmetric) ``intersection'' matrices 
\begin{equation}
\begin{aligned}
  h_{st} \overline{\alpha}^s _k \wedge \alpha^t _l = c^i _{kl} \eta_i \ , \qquad
h_{st} \alpha_k ^s \wedge \alpha_l ^t = e_{kl} \overline{\Omega} \ .
\end{aligned}
\end{equation}

The scalar target manifold  is a fibration of the
bundle moduli $\xi$ 
and the  charged scalars $C$ 
over the  $K3$ moduli space $\mathcal M$ given in \eqref{K3mod2}. 
Supersymmetry imposes that this scalar manifold is quaternionic-K\"ahler 
which, however, we did not verify explicitly.
In appendix~\ref{limit} we show that our results are consistent with the orbifold limit $T^4/\mathbb{Z}_3$ (with standard embedding).
The scalars of the truncated spectrum corresponding to the untwisted sector span the quaternionic-K\"ahler (and simultaneously K\"ahler) manifold
\begin{equation}
 \frac{SU(2,2+56)}{U(1)\times SU(2)\times SU(2+56)} \ .
\end{equation}

We now turn to the scalar potential
which consists of all terms descending from \eqref{hetaction10} with space-time indices tangent to $K3$. Since $K3$ is Ricci-flat the Gauss-Bonnet term \eqref{GB} reduces 
to the square of the curvature 2-form. Moreover, since the curvature is anti-selfdual
for all metric deformations, the term gives a constant topological contribution equal to the Euler number of $K3$
\begin{equation}
 -\tfrac12 \int \limits_{K3} \mbox{tr} ({\tilde{\mathcal R}} \wedge \star {\tilde{\mathcal R}}) = \tfrac12 \int \limits_{K3} \mbox{tr} ({\mathcal R} \wedge {\mathcal R}) = 24 \ .
\end{equation}
Together with the contribution from the Yang-Mills field strength we obtain 
\begin{equation}
V = -\tfrac{\alpha'}{2 \mathcal V} e^{\phi} \Bigl( \int \mbox{tr}  F_{\bar 2} \wedge \star F_{\bar 2} + 48 \Bigr) \ .
\end{equation}
Dividing into background and fluctuations $F_{\bar 2}= \mathcal F+ \Fl_{\bar 2}$ we arrive at
\begin{equation}
V = - \tfrac{\alpha'}{\mathcal V} e^{\phi} \Bigl( -\tfrac{1}{2}\int \mbox{tr} ({\mathcal F}
\wedge {\mathcal F}) + 24 + \tfrac{1}{2} \int \mbox{tr}( \Fl_{\bar
  2} \wedge \star \Fl_{\bar 2}) \Bigr)\ .
\end{equation}
The first two terms vanish due to the tadpole condition \eqref{BI2}
while the third can be decomposed into selfdual and
anti-selfdual parts.
The tadpole condition additionally constrains 
\begin{equation}\label{tad3}
0 = \int\mbox{tr} ( \Fl_{\bar{2}} \wedge  \Fl_{\bar{2}} ) = \int\mbox{tr} ( \Fl_{\bar{2}+} \wedge  \Fl_{\bar{2}+} )
+\int \mbox{tr} ( \Fl_{\bar{2}-} \wedge \Fl_{\bar{2}-} ) \ ,
\end{equation}
since continuous fluctuations cannot change a topological invariant.
Therefore we can express the potential entirely in terms of the selfdual part $\Fl_{\bar{2}+}$ to obtain
\begin{equation}\begin{aligned}\label{Vym}
V &= -\tfrac{\alpha'}{2 \mathcal V} e^{\phi} \int  \mbox{tr}( \Fl_{\bar 2} \wedge \star \Fl_{\bar 2}) \\
&= -\tfrac{\alpha'}{2 \mathcal V} e^{\phi} \int \mbox{tr} \left( \Fl_{\bar{2}+} \wedge \Fl_{\bar{2}+} - \Fl_{\bar{2}-} \wedge  \Fl_{\bar{2}-} \right) \\
&= - \tfrac{\alpha'}{\mathcal V} e^{\phi} \int \mbox{tr} ( \Fl_{\bar{2}+} \wedge \Fl_{\bar{2}+} ) \ .
\end{aligned}\end{equation}
This is positive definite since for antihermitean generators the trace
gives a negative Killing form.

One thus has to compute the selfdual components $\Fl_{\bar{2}+} ^{({\bf
    1}, {\bf 3})}, \Fl_{\bar{2}+} ^{({\bf 133}, {\bf 1})}$ of
the terms
given in \eqref{problem} and \eqref{Dse}. In appendix \ref{potSE} we show
that $\Fl_{\bar{2} +} ^{({\bf 1}, {\bf 3})}$
vanishes, due to the nontriviality of the adjoint $\mbox{End} \ {\mathcal T}_{K3}$-bundle. 
This is crucial for consistency with 6D supergravity, since $D$-terms
necessarily are valued in the adjoint of the unbroken gauge group.
On the other hand, the selfdual part of \eqref{Dse} reads
\begin{equation}\label{Dse+}
 \Fl_{\bar{2}+} ^{({\bf 133},{\bf 1})} \equiv \Fl_{\bar{2}+} ^{a} = \begin{pmatrix} \overline{C}^x _i \\ C^x _i \end{pmatrix}^T  \begin{pmatrix} -i \sqrt{2 \mathcal V} G_{ij}
 J_3 & \tfrac12 \tilde{\rho}_{ij} \Omega \\ \tfrac12 \tilde{\rho}_{ij} \overline{\Omega} & i\sqrt{2 \mathcal V} G_{ij} J_3 \end{pmatrix} 
(\tau^a)_{xy} \begin{pmatrix} C^y _j \\ \overline{C}^y _j \end{pmatrix} \ 
\end{equation}
(see \eqref{Dse++}). Here $\tilde{\rho}_{ij} = \gamma_i \gamma_j \rho_{ij}$ is the rescaled $K3$-intersection matrix \eqref{insect}, restricted to $H^{1,1} (K3, \mathbb{R})$ and $G_{ij}$ is the same coupling as in \eqref{smetric}. 
The $D$-term is identified by expanding
\begin{equation}
 \Fl_{\bar{2}+} ^{a} 
= \left(\int \Fl_{\bar{2}+} ^a \wedge J_s \right) J_s\ ,
\end{equation}
Inserting this into \eqref{Vym} we arrive at
\begin{equation}\label{D2}
\begin{aligned}
 V &= -\tfrac{\alpha'}{\mathcal V} e^{\phi} \int J_s \wedge J_t \left(\int \Fl_{\bar{2}+} ^a \wedge J_s \right) \left( \int \Fl_{\bar{2}+} ^a \wedge J_t \right) \\
&= -\tfrac{\alpha'}{2 \mathcal V} e^{\phi} \mbox{tr} (\sigma^{(s)} \sigma^{(t)})
\left(\int \Fl_{\bar{2}+} ^a \wedge J_s \right) \left( \int
   \Fl_{\bar{2}+} ^a \wedge J_t \right)  \ .
\end{aligned}
\end{equation}
Comparing with the generic scalar potential \eqref{V6} yields
\begin{equation}\label{D}
(D^{a})^A _{\ B} = \tfrac{1}{\sqrt{2 \mathcal V}} (\sigma^{(s)})^A _{\ B} \int \Fl_{\bar{2}+} ^a \wedge J_s \ .
\end{equation}
Hence, the standard embedding on $K3$ leads to a quartic $D$-term potential for the charged scalars in consistency with the generic 6D supergravity. 
If there exist $D$-flat directions the moduli space of vacua has a Higgs branch, where the gauge group is broken further.

Finally, we identify the $\mathfrak{su}(2)_R$-valued connection 1-form $\A$ on the charged scalar field space, defined in \eqref{Dt}.
Separating the Killing vectors $K_i ^{xa} = (\tau^a)_y ^{\ x} C_i
^y$
in the $D$-term \eqref{D} yields
\begin{equation}
 ({\A}^x _j)^A_{\ B} =  \begin{pmatrix}  i G_{ij} (-\overline{C}^x _i, C^x _i) & \tfrac{1}{\sqrt{2 \mathcal V}}\tilde{\rho}_{ij} (0,\overline{C}^x _i) \\
 \tfrac{1}{\sqrt{2 \mathcal V}}\tilde{\rho}_{ij} (C^x _i, 0) & i G_{ij} (\overline{C}^x _i, -C^x _i) \end{pmatrix}^A _{\ B} \ .
\end{equation}
The corresponding curvature tensor is nonvanishing. 

\subsection{Deviation from the standard embedding}
\label{dev}
Before we continue let us briefly discuss the scalar potential
for deviations from the standard embedding.
A first generalization is to drop the condition ${\mathcal F} =
{\mathcal R}$ but keep the anti-selfduality of ${\mathcal F}$.
This is automatically satisfied for any 
instanton configuration. 
In this class the scalar potential for the $K3$ moduli is trivially zero.
The second generalization is to consider an arbitrary Yang-Mills bundle.
Under metric deformations the curvature 2-form of $K3$ stays anti-selfdual, but the Yang-Mills curvature generically loses this property. In this case the selfdual part contributes an additional term to the 
scalar potential given by
\begin{equation}\label{shift2}
\begin{aligned}
 V_6 &\sim -\tfrac{1}{2} \int \mbox{tr} ({\mathcal F} \wedge \star {\mathcal F}) + \tfrac{1}{2} \int \mbox{tr} ({\mathcal R} \wedge {\mathcal R}) \\
&= - \tfrac{1}{2} \int \mbox{tr} ({\mathcal F} \wedge \star {\mathcal F}) -\tfrac{1}{2} \int \mbox{tr} ({\mathcal F} \wedge {\mathcal F}) \\
&= - \int \mbox{tr} ({\mathcal F}_+ \wedge \star {\mathcal F}_+) \ .
\end{aligned}
\end{equation}
This term is positive definite, because the Killing form is negative on antihermitean generators. 
There are two ways how the system can go back to the minimum of the potential. Either $\mathcal F$ is dynamically driven to a new ASD ground state or the $K3$ metric deforms 
in such a way that $\mathcal F$ becomes anti-selfdual again. It follows that for a fixed $\mathcal F$ only metric deformations which preserve the ASD condition are true moduli, 
while the others generate a potential like \eqref{shift2}. 
In the next section we will consider Yang-Mills fluxes which are rigid backgrounds, fixed by a quantization condition. In particular they cannot deform dynamically to different 
ASD ground states. This will stabilize some of the $K3$ metric moduli.

\section{Line bundles on $K3$}\label{LB}
In this section we look for solutions of the tadpole condition
different from the standard embedding, i.e.\ backgrounds which only
satisfy 
the integrated equation \eqref{BI} in terms of characteristic classes. 
Strictly speaking, this is not possible with ${\mathcal H} \equiv 0$ in this background. One has to include torsion into the internal geometry and the proper 
back reaction is given by the Strominger equations. For six internal dimensions one loses the Calabi-Yau property or even more structure, but for $K3$ the torsion can be 
completely absorbed in a conformal factor of the metric \cite{Strominger:1986uh}. 

In the following we consider $K3$ compactifications with line bundles,
where the tadpole condition is solved by assigning ${\mathcal F}$ to be
the curvature of one (or several) principal $U(1)$ bundle(s) \cite{Green:1984bx,Aldazabal:1996du, Louis:2001uy, Andreas:2004ja, Blumenhagen:2005ga,Blumenhagen:2006ux,Honecker:2006dt, Honecker:2006qz, Kumar:2009ae,Kumar:2009zc}. For one
line bundle $L$
inside one $E_8$ factor we then have
\begin{equation}
 E_8 \longrightarrow G \times \langle U(1) \rangle \ ,
\end{equation}
which implies the following
decomposition of the adjoint representation
\begin{equation}\label{dec3}
 {\bf 248} \longrightarrow
({\mathfrak g}, {\bf 1}_0) \oplus ({\bf 1}, {\bf 1}_0)\
 \bigoplus \limits_i \ \left(({\bf R}_i,{\bf 1}_{q_i}) \oplus ({\overline{\bf R}}_i,{\bf 1}_{-q_i})\right)  \ ,
\end{equation}
where ${\mathfrak g}$ is the adjoint representation of $G$
while the second term includes ${\bf 1}_0$ as the adjoint representation
of $U(1)$. The ${\bf R}_i$ are model dependent representations of $G$
and ${\bf 1}_{q_i}$ are representations of $U(1)$ with charge
$q_i$. The right entries define associated vector bundles
$E_{{\bf 1}_q}$ which are  tensor products of the 
line bundle $L$ with charge $q$:
\begin{equation}
 E_{{\bf 1}_q} = L^q = L \otimes ... \otimes L \ .
\end{equation}
Negative charges correspond to the dual bundle, $L^{-1} = L^*$, and $L^0 = {\mathcal O}$ is the trivial bundle. 

Applying the deformation theory of gauge connections to this setup
(for more details see appendix \ref{line})
yields the multiplicities of the corresponding massless fields.
Specifically one finds
\begin{equation}\label{chiline}
h^{0,1} (L^q)= 
-2 - q^2 ch_2(L) \ ,
\end{equation}
where  $ch_2(L)= -\tfrac{1}{2} \int 
\mbox{tr}{\mathcal F} \wedge {\mathcal F}$ is the second Chern-character.
Moreover, no bundle moduli exist, as $\mbox{End} \ L^q$ is the
trivial line bundle with $H^{0,1}(\mbox{End} \ L^q) = 0$.
%
%
Since the only nonvanishing Chern class is
$c_1(L) = i\, \mbox{tr} {\mathcal F} \in H^{1,1}(K3, \mathbb{Z})$, nontrivial line bundles are equivalent to integral, Abelian Yang-Mills fluxes.\footnote{There exist no 
Abelian local instantons on $K3$ because in 4D these are characterized by the winding number of the mapping $S^3 \mapsto U(1)$, however $\pi_3(U(1)) = 0$.} Therefore, to specify a 
line bundle, one chooses a vector $\flux$ in the Cartan subalgebra $E_8 \times E_8$ and an integral linear combination of the 2-cycles of $K3$.\footnote{The specific choice of 2-cycles can be motivated 
by making contact with heterotic orbifold models which arise as singular
limits of $K3$ with shrinking 2-cycles \cite{Honecker:2006qz,Nibbelink:2007rd}.} 
$\flux$ determines the group theoretical 
embedding and the unbroken gauge group while the 2-cycles determine the location of the flux 
\begin{equation}\label{flux}
i{\mathcal F} = \flux \otimes m^I \eta_I \ , \qquad I=1,...,22 \ ,
\end{equation}
with $\eta_I$ being an integral basis of $H^2(K3,\mathbb{Z})$. 
The flux satisfies the quantization condition
\begin{equation}\label{quant}
 i \int \limits_{\Gamma^I} \mbox{tr}{\mathcal F} = - \lVert \flux \rVert \ m^I \ \ \in \mathbb{Z} \ ,
\end{equation}
for all integral 2-cycles $\Gamma_I \in H_2(K3, \mathbb{Z})$. Here
$\lVert \flux \rVert$ is the Euclidean norm in the Cartan subalgebra of $E_8$. 
For a given $K3$-metric a supersymmetry 
preserving background must in addition satisfy the ASD condition
${\mathcal F} \in H^{1,1}_- (K3, \mathbb{Z})$, which is a restriction
on the $K3$ metric as we already said in section~\ref{dev}.

We can extend the construction to several line bundles, each with field strength
\begin{equation}
i{\mathcal F}^n = \flux^n \otimes m^{In} \eta_I \ .
\end{equation}
Since $E_8 \times E_8$ has rank 16, there are at most 16 independent 
line bundles available. For the tadpole condition we must have
\begin{equation}\label{tadpole}
\begin{aligned}
 24 &= \tfrac{1}{2} \int \mbox{tr} ({\mathcal F} \wedge {\mathcal F}) 
 &= - \tfrac{1}{2} (\flux^n\cdot \flux^m) \ m^{In} m^{Jm} \rho_{IJ} \ .
\end{aligned}
\end{equation}
Here $\cdot$ is the Euclidean scalar product in the Cartan subalgebra and $\rho_{IJ}$ is the 2-cycle intersection matrix \eqref{insect} of $K3$.

\subsection{Reduction of the Yang-Mills sector}
Using the results from appendix \ref{line}, the Kaluza-Klein expansion of the gauge potential reads
\begin{equation}\label{Aline}
 \dA_1 = \V^{\mathfrak g} + \V^{\bf 1} \ , \qquad 
 \dA_{\bar 1}\ = \ \sum_{i} \ (C_{k_i} ^{{\bf R}_i} \omega_{k_i} ^{q_i} + \overline{C}_{k_i} ^{\overline{\bf R}_i} \overline{\omega}_{k_i} ^{-q_i})  + (\overline{D}_{k_i} ^{\overline{\bf R}_i} \overline{\varpi}_{k_i} ^{-q_i} + D_{k_i} ^{{\bf R}_i} \varpi_{k_i} ^{q_i}) \ .
\end{equation}
Here $V^{\mathfrak g}$ is the 6D gauge potential in the adjoint representation of $G$. For one line bundle, we have additionally the Abelian gauge potential $V^{\bf 1}$. 
For $q_i \neq 0$ the representations in \eqref{dec3} are complex and always occur pairwise, 
with corresponding charged scalars $C_{k_i}$ and $\overline{D}_{k_i}$, respectively. 
Their four real degrees of freedom align in one 
hypermultiplet in the representation
${\bf R}_i \oplus \overline{\bf R}_i$. The zero modes belong to
\begin{equation}
\begin{aligned}
 \omega_{k_i} ^{q_i} \in H^{0,1} (L^{q_i}) \ &, \ \ \ \ 
 \overline{\omega}_{k_i} ^{-q_i} \in H^{1,0} (L^{-q_i}) \ , \\
 {\varpi}_{k_i} ^{q_i} \in H^{1,0} (L^{q_i}) \ &, \ \ \ \ 
 \overline{\varpi}_{k_i} ^{-q_i} \in H^{0,1} (L^{-q_i}) \ ,
\end{aligned}
\end{equation}
with multiplicities $k_i = 1,..., h^{0,1}(L^{q_i})$. For notational 
simplicity we define doublets of the charged scalars as
\begin{equation}
 \Phi^{{\bf R}_i} _{k_i} := (C^{{\bf R}_i} _{k_i}, D^{{\bf R}_i} _{k_i}) \ .
\end{equation}
 
From \eqref{Aline} we derive the Kaluza-Klein expansion of the field strength
\begin{equation}
 \Fl =  \Fl_2 ^{\bf 1} + \Fl_2 ^{\bf \mathfrak g}  + \sum \limits_i (\Fl_{1,\bar{1}} ^{{\bf R}_i} + \overline{\Fl}_{1,\bar{1}} ^{\overline{\bf R}_i}) +  \Fl_{\bar 2} \ .
\end{equation}
Here $\Fl_2^{\bf 1} = d\V^{\bf 1}$ and $\Fl_2 ^{\bf \mathfrak g} = 
d\V^{\mathfrak g} + \tfrac12[V^{\mathfrak g}, V^{\mathfrak g}]$ are the 6D field strengths. 
The terms with one external and one internal tangent index give rise to gauge covariant derivatives of the charged scalars,
\begin{equation}\label{6cov}
\begin{aligned}
 f_{1,\overline{1}} ^{{\bf R}_i} &= {\mathcal D}\Phi^{{\bf R}_i} _{k_i} \wedge \omega^{q_i} _{k_i}, \qquad
 {\mathcal D}\Phi^{{\bf R}_i} = d\Phi^{{\bf R}_i} -q_i V^{\bf 1} \Phi^{{\bf R}_i} - V^a (\tau_a \Phi)^{{\bf R}_i} \ .
\end{aligned}
\end{equation}
Using the zero mode property $d_{\mathcal A} \omega_{k_i} ^{q_i} = d_{\mathcal A} \varpi_{k_i} ^{q_i} = 0$, the internal fluctuation is given by the commutator
\begin{equation}\label{comm3}
 \Fl_{\bar 2} = \tfrac12 \sum \limits_{i,j} \left[\dA_{\bar 1} ^{({\bf R}_i, {\bf 1}_{q_i})}, \dA_{\bar 1} ^{({\bf R}_j, {\bf 1}_{q_j})}\right] \ .
\end{equation}
Depending on the surviving gauge group $G$, several representations
can arise
in \eqref{comm3}.
For $i=j$ the commutator generates the adjoint representations of the unbroken gauge group $G \times U(1)$
\begin{equation}
({\bf R}_i, {\bf 1}_{q_i}) \otimes (\overline{\bf R}_i, {\bf 1}_{-q_i}) = ({\mathfrak g}, {\bf 1}_0) \oplus ({\bf 1}, {\bf 1}_0) \oplus ... \ .
\end{equation}
It results in field strength fluctuations of the form
\begin{equation}\label{Dline1}
\Fl_{\bar 2} ^{\mathfrak g} = \sum \limits_i \begin{pmatrix} \overline{C}^{\overline{\bf R}_i} \\ \overline{D}^{\overline{\bf R}_i} \end{pmatrix}^T \begin{pmatrix} \overline{\omega}^{-q_i} \wedge \omega^{q_i} & \overline{\omega}^{-q_i} \wedge \varpi^{q_i} \\ \overline{\varpi}^{-q_i} \wedge \omega^{q_i} & \overline{\varpi}^{-q_i} \wedge \varpi^{q_i} \end{pmatrix} (\tau^a) \begin{pmatrix} C^{{\bf R}_i} \\ D^{{\bf R}_i} \end{pmatrix} \ ,
\end{equation}
\begin{equation}\label{Dline2}
\Fl_{\bar 2} ^{\bf 1} = \sum \limits_i q_i \begin{pmatrix}
   \overline{C}^{\overline{\bf R}_i} \\ \overline{D}^{\overline{\bf
       R}_i} \end{pmatrix}^T \begin{pmatrix} \overline{\omega}^{-q_i}
   \wedge \omega^{q_i} & \overline{\omega}^{-q_i} \wedge \varpi^{q_i}
   \\ \overline{\varpi}^{-q_i} \wedge \omega^{q_i} &
   \overline{\varpi}^{-q_i} \wedge \varpi^{q_i} \end{pmatrix} (\mathbb{I}) \begin{pmatrix} C^{{\bf R}_i} \\ D^{{\bf R}_i} \end{pmatrix} \ ,
\end{equation}
where we suppressed the multiplicity indices. $\tau^a$ are the $\mathfrak g$-generators in the appropriate representation ${\bf R}_i$. 
The products of zero modes belong to $H^2 (L^{q_i} \otimes L^{-q_i}) = H^2(K3, \mathbb{R})$. 
Other representations can occur if the adjoint decomposition allows for other tensor products. 
Let us illustrate this with an explicit example: There exists a Cartan generator for the line bundle \cite{Green:1984bx, Honecker:2006qz} that breaks
\begin{equation}
\begin{aligned}
 E_8 &\longrightarrow SO(14) \times U(1): \\
{\bf 248} &\longrightarrow {\bf 91}_0 \oplus {\bf 1}_0 \oplus ({\bf 64}_1 \oplus \overline{\bf 64}_{-1}) \oplus ({\bf 14}_2 \oplus \overline{\bf 14}_{-2}),
\end{aligned}
\end{equation}
where $\bf 64$ is the Weyl-spinor of $SO(14)$. Then the commutator \eqref{comm3} realizes the tensor products
\begin{equation}\label{tproducts}
 \begin{aligned}
  {\bf 64}_1 \otimes {\bf 64}_1 &= {\bf 14}_2 \oplus . . . , \\
  \overline{\bf 64}_{-1} \otimes \overline{\bf 64}_{-1} &= \overline{\bf 14}_{-2} \oplus . . . , \\
  {\bf 64}_1 \otimes \overline{\bf 14}_{-2} &= \overline{\bf 64}_{-1} \oplus . . . , \\
  \overline{\bf 64}_{-1} \otimes {\bf 14}_2 &= {\bf 64}_1 \oplus . . .
  \ .
 \end{aligned}
\end{equation}
The first two tensor products
generate a field strength fluctuation of the form
\begin{equation}\label{problem2}
\begin{aligned}
\Fl_{\bar 2} ^{{\bf 14}_2 \oplus {\bf 14}_{-2}} = &\begin{pmatrix} C^u \\ D^u \end{pmatrix}^T \begin{pmatrix} \omega^1 \wedge \omega^1 & \omega^1 \wedge \varpi^1 \\ \varpi^1 \wedge \omega^1 & \varpi^1 \wedge \varpi^1 \end{pmatrix} (\sigma^x)_{uv} \begin{pmatrix} C^v \\ D^v \end{pmatrix} \\
&+ \begin{pmatrix} \overline{C}_u \\ \overline{D}_u \end{pmatrix}^T \begin{pmatrix} \overline{\omega}^{-1} \wedge \overline{\omega}^{-1} & \overline{\omega}^{-1} \wedge \overline{\varpi}^{-1} \\ \overline{\varpi}^{-1} \wedge \overline{\omega}^{-1} & \overline{\varpi}^{-1} \wedge \overline{\varpi}^{-1} \end{pmatrix} (\sigma_x)^{uv} \begin{pmatrix} \overline{C}_v \\ \overline{D}^v \end{pmatrix} \ ,
\end{aligned}
\end{equation}
where we again suppressed the multiplicity indices. The products of zero
modes belong to $H^{1,1}(L^2 \oplus L^{-2})$. The latter two tensor
products in \eqref{tproducts} yield an analogous term 
$\Fl_{\bar 2} ^{{\bf 64}_1 \oplus {\overline{\bf 64}}_{-1}}$. Together we have for this example
\begin{equation}\label{deltaF}
 \Fl_{\bar 2} \ = \ \Fl^{{\bf 91}_0} _{\bar 2} \ + \ \Fl^{{\bf 1}_0} _{\bar 2} \ + \ \Fl^{{\bf 14}_2 \oplus {\bf 14}_{-2}} _{\bar 2} \ + \ \Fl^{{\bf 64}_1 \oplus {\overline{\bf 64}}_{-1}} _{\bar 2}\ .
\end{equation}

\subsection{Reduction of the Kalb-Ramond sector}\label{KR2}
The reduction is essentially the same as in section \ref{KR1}, so we only present the new features. 
The coupling between the $b$-scalars and the charged scalars again arises from the $\omega_{1, \bar 2} ^{YM}$ component of the Chern-Simons 3-form.
But due to the Abelian character of the flux the nonvanishing terms are
\begin{equation}\label{CSint2}
 \omega_{1, \bar 2} ^{YM} = \sum \limits_i \mbox{tr} \Bigl(\Fl_{1,\bar 1} ^{{\bf R}_i} \wedge \dA_{\bar 1} ^{({\bf R}_i, {\bf 1}_{q_i})} \Bigr)
 + \mbox{tr} \Bigl( {\mathcal F}^{\bf 1} \wedge \dA_1 ^{\bf 1} \Bigr) \ .
\end{equation}
Compared to \eqref{CSint} we see that the second term in \eqref{CSint2}
vanishes in the standard embedding (as well as for any non-Abelian
gauge bundle) since in that case there cannot be a 6D vector in the same
representation as the background field strength ${\mathcal F}$.
The first term in \eqref{CSint2} generates the skew-symmetric $\bar{\Phi} \overleftrightarrow{\mathcal D} \Phi$ couplings and the second term affinely gauges the $b$-scalars under the unbroken $U(1)$.
Using the expansion ${\mathcal F} = -i \flux m^I \eta_I$ we get
\begin{equation}
 dB_{1, \bar 2} + \alpha'  \omega_{1, \bar 2} ^{YM} = \left(db^I - \alpha' V^{\bf 1} \lVert \flux \rVert^2 m^I\right) \eta_I + \alpha' \sum \limits_i \mbox{tr} \Bigl(\overline{\Phi}^{\overline{\bf R}_i} 
\overleftrightarrow{\mathcal D} \Phi^{{\bf R}_i}\Bigr) \ ,
\end{equation}
 where 
\begin{equation}
\begin{aligned}
 \overline{\Phi}^{\overline{\bf R}_i} \overleftrightarrow{\mathcal D} \Phi^{{\bf R}_i} = \ &\frac{1}{2} \begin{pmatrix} \overline{C}^{\overline{\bf R}_i} _{k_i} \\                                                                                     
 \overline{D}^{\overline{\bf R}_i} _{k_i} \end{pmatrix} 
\begin{pmatrix} \overline{\omega}^{-q_i} _{k_i} \wedge \omega^{q_i} _{l_i} & \overline{\omega}^{-q_i} _{k_i} \wedge \varpi^{q_i} _{l_i} \\ 
\overline{\varpi}^{-q_i} _{k_i} \wedge \omega^{q_i} _{l_i} & \overline{\varpi}^{-q_i} _{k_i} \wedge \varpi^{q_i} _{l_i} \end{pmatrix}
\begin{pmatrix} {\mathcal D} C^{{\bf R}_i} _{l_i} \\ {\mathcal D} D^{{\bf R}_i} _{l_i} \end{pmatrix} \\ 
&- \frac{1}{2} \begin{pmatrix} {\mathcal D}\overline{C}^{\overline{\bf R}_i} _{k_i} \\                                                                                     
 {\mathcal D}\overline{D}^{\overline{\bf R}_i} _{k_i} \end{pmatrix} 
\begin{pmatrix} \overline{\omega}^{-q_i} _{k_i} \wedge \omega^{q_i} _{l_i} & \overline{\omega}^{-q_i} _{k_i} \wedge \varpi^{q_i} _{l_i} \\ 
\overline{\varpi}^{-q_i} _{k_i} \wedge \omega^{q_i} _{l_i} & \overline{\varpi}^{-q_i} _{k_i} \wedge \varpi^{q_i} _{l_i} \end{pmatrix}
\begin{pmatrix} C^{{\bf R}_i} _{l_i} \\ D^{{\bf R}_i} _{l_i} \end{pmatrix} \ .
\end{aligned}
\end{equation}

\subsection{6D Effective action}
Let us now turn to the effective action combining the previous results
\begin{equation}\label{Kinetic6Dterms2}
\begin{aligned}
 {\mathcal L}_6 = & \ \tfrac12 R * 1 \ - \ \tfrac16 e^{-2\phi}  H \wedge * H \ + \ \tfrac{\alpha'}{2} e^{-\phi} \mbox{tr} F^{\mathfrak g} \wedge * F^{\mathfrak g} 
\ - \ \tfrac{\alpha'}{2} e^{-\phi} \lVert \flux \rVert^2 F^{\bf 1} \wedge * F^{\bf 1} \\
 &+ \tfrac92 d\phi \wedge * d\phi \ + \ \tfrac14 h_{IJ} dt_s ^I \wedge * dt_s ^J \ - \ \tfrac{1}{8{\mathcal V}^2} d{\mathcal V} \wedge * d{\mathcal V} \\
 &- \ \alpha' \sum \limits_i G_{k_i l_i} ^{{\bf R}_i} 
{\rm tr} \big({\mathcal D}\overline{\Phi}^{\overline{\bf R}_i} _{k_i} \wedge *
{\mathcal D}\Phi^{{\bf R}_i}_{l_i} \big)
\ - \ \tfrac1{6\mathcal{V}}\, g_{IJ} {\mathcal D} b^I \wedge * {\mathcal D} b^J  - V *1 \ .
\end{aligned}
\end{equation}
$F^{\mathfrak g}$ is the Yang-Mills field strength of the semi-simple part of the unbroken gauge group and $F^{\bf 1}$ is 
the field strength of the unbroken $U(1)$ corresponding to the line bundle. The derivatives of the scalars read
\begin{equation}\label{db2}
\begin{aligned}
 {\mathcal D}\Phi^{{\bf R}_i} &= d\Phi^{{\bf R}_i} -q_i V^{\bf 1} \Phi^{{\bf R}_i} - V^a (\tau_a \Phi)^{{\bf R}_i} \ , \\
{\mathcal D} b^I &= db^I - \alpha' V^{\bf 1} \lVert \flux \rVert^2  m^I + \alpha' \rho^{IJ} \mbox{tr}
\Bigl(\overline{\Phi}^{\overline{\bf R}_i} _{k_i}  (N_{J k_i l_i} ^{q_i}) \overleftrightarrow{\mathcal D} \Phi^{{\bf R}_i} _{l_i}\Bigr) \ .
\end{aligned}
\end{equation}
We see that the scalars $\Phi^{{\bf R}_i}$ are linearly gauged under the entire unbroken gauge group. The $b$-scalars are affinely gauged under the unbroken $U(1)$ due to the flux of the line bundle, 
with charges given by the flux vector $m^I$. The $2\times 2$ coupling matrix $(N_{J k_i l_i} ^{q_i})$ is given by
\begin{equation}
 (N_{J k_i l_i} ^{q_i}) = \int \eta_J \wedge \begin{pmatrix} \overline{\omega}^{-q_i} _{k_i} \wedge \omega^{q_i} _{l_i} & \overline{\omega}^{-q_i} _{k_i} \wedge \varpi^{q_i} _{l_i} \\ 
\overline{\varpi}^{-q_i} _{k_i} \wedge \omega^{q_i} _{l_i} & \overline{\varpi}^{-q_i} _{k_i} \wedge \varpi^{q_i} _{l_i} \end{pmatrix} \ .
\end{equation}
The scalar metrics read
\begin{equation}\label{skinetic}
\begin{aligned}
 g_{IJ} &= \int \eta_I \wedge \star \eta_J \ , \\
 (G^{{\bf R}_i} _{k_i l_i}) &= {\mathcal V}^{-1} \int \begin{pmatrix} \omega^{q_i} _{k_i} \wedge \star \overline{\omega}^{-q_i} _{l_i} & 0 \\ 
 0 & \varpi^{q_i} _{k_i} \wedge \star \overline{\varpi}^{-q_i} _{l_i} \end{pmatrix} \ ,
\end{aligned}
\end{equation}
so the latter is diagonal in the $C^{{\bf R}_i}$ and $D^{{\bf R}_i}$ fields.

We now turn to the scalar potential. By the same argument as given in 
\eqref{Vym} for the standard embedding, only the selfdual parts of the field strength fluctuations \eqref{deltaF} 
contribute to the potential. It is shown in appendix \ref{line} that the selfdual parts vanish for all terms which are not in the adjoint representation of the unbroken gauge group  
\begin{equation}
 \Fl^{{\bf R}_i \oplus {\bf R}_i} _{{\bar 2}+} = 0 \ .
\end{equation}
On the other hand the selfdual parts of \eqref{Dline1} and
\eqref{Dline2} take the form
\begin{equation}\label{Dline3}
\Fl_{\bar{2}+} ^{\mathfrak g} = \sum \limits_i
\overline{\Phi}^{\overline{\bf R}_i} _{k_i} (\U_{k_i l_i})
(\tau^a \Phi)^{{\bf R}_i} _{l_i}\ ,\qquad
\Fl_{\bar{2}+} ^{\bf 1} =\sum \limits_i q_i
\overline{\Phi}^{\overline{\bf R}_i} _{k_i} (\U_{k_i l_i})
\Phi^{{\bf R}_i} _{l_i}\ ,
\end{equation}
where
\begin{equation}\label{Dline4}
\U_{k_i l_i} = \begin{pmatrix} \tfrac{i}{2} G_{k_i l_i} ^C J & \tfrac{1}{2} c_{k_i l_i} \Omega \\[1ex]
\tfrac{1}{2}\overline{c}_{k_i l_i} \overline{\Omega} & \tfrac{i}{2}
G_{k_i l_i} ^D J \end{pmatrix}\ .
\end{equation}
Note that $a$ is used for the adjoint $\mathfrak g$ index and that the
matrix $\U$ depends on the representation ${\bf R}_i$. As in the standard embedding we find on the diagonal the scalar metrics $G_{k_i l_i} ^C$ and $G_{k_i l_i} ^D$, 
which are the two matrix elements of \eqref{skinetic}. 
In the off-diagonal elements we find a generalized ``intersection matrix''
\begin{equation}
c_{k_i l_i} = \int \overline{\omega}_{k_i} ^{-q_i} \wedge \varpi_{l_i} ^{q_i} \wedge \overline{\Omega} \ .
\end{equation}
Identifying the Killing vectors
\begin{equation}
 K_{k_i} ^a = (\tau^a \Phi_{k_i})^{{\bf R}_i} \ , \qquad K^{\bf 1} _{k_i} = q_i \Phi_{k_i}^{{\bf R}_i} \ , \qquad K^{I \bf 1} = \lVert \flux \rVert^2 m^I \ ,
\end{equation}
we see that the terms \eqref{Dline3}, \eqref{Dline4} generate $D$-terms in the 6D potential. The third Killing vector corresponds to the gauge flux, 
whose selfdual component appears as a Fayet-Iliopoulos term in the Abelian $D$-term
\begin{equation}\label{FI}
 V = - \tfrac{\alpha'}{\mathcal V} e^{\phi} \int  \mbox{tr} \left(({\mathcal F}_+ + \Fl_{\bar{2}+} ^{\bf 1}) \wedge \star ({\mathcal F}_+ + \Fl_{\bar{2}+} ^{\bf 1})\right) - \tfrac{\alpha'}{\mathcal V} e^{\phi}
\int  \mbox{tr} \left( \Fl_{\bar{2}+} ^{\bf \mathfrak{g}} \wedge \star \Fl_{\bar{2}+} ^{\bf \mathfrak{g}}\right) \ .
\end{equation}
Similar to the analysis in \eqref{D2} and \eqref{D},
the individual $D$-terms can be extracted from \eqref{FI} 
by the ($K3$ metric dependent) expansion 
\begin{equation}
\begin{aligned}
 (D^a)^A _{\ B} &= \tfrac{1}{\sqrt 2} \int \Fl_{\bar{2}+} ^a \wedge J_s \otimes (\sigma^{(s)})^A _{\ B} \ , \\
 (D^{\bf 1})^A _{\ B} &= \tfrac{1}{\sqrt 2} \int ({\mathcal F}_+ + \Fl_{\bar{2}+} ^{\bf 1}) \wedge J_s \otimes (\sigma^{(s)})^A _{\ B} \ .
\end{aligned}
\end{equation}

The generalization of the above results to several line bundles is straightforward. The $b$-scalars are then gauged under all Abelian factors $U(1)_m$ with charges proportional to the 
flux vectors $m^{In}$. For line bundles which are not orthogonal, $X^n \cdot X^m \neq 0$, kinetic mixing of the different $F^{{\bf 1}_m}$ field strengths occurs
\begin{equation}
 {\mathcal L}_6 ^{kin} \sim - \tfrac{\alpha'}{2} e^{-\phi} \sum \limits_{m,n} (X^m \cdot X^n) F^{{\bf 1}_m} \wedge * F^{{\bf 1}_n} \ .
\end{equation}
Also the 6D $H$-field may contain mixed Abelian Chern-Simons couplings
(see \eqref{mixed}). 
The scalar potential takes the form
\begin{equation}
 V = \tfrac{\alpha'}{\mathcal V} e^{\phi} (X^n \cdot X^m) \int ({\mathcal F}_+ ^n + f_{{\bar 2}+} ^{{\bf 1}_n}) \wedge \star ({\mathcal F}_+ ^m + f_{{\bar 2}+} ^{{\bf 1}_m})
- \tfrac{\alpha'}{\mathcal V} e^{\phi} \int  \mbox{tr} \left( \Fl_{\bar{2}+} ^{\bf \mathfrak{g}} \wedge \star \Fl_{\bar{2}+} ^{\bf \mathfrak{g}}\right) \ .
\end{equation}
Here $f_{{\bar 2}+} ^{{\bf 1}_n}$ is the direct generalization of \eqref{Dline2} containing all charged matter fields charged under $U(1)_n$.
The explicit form of the scalar potential reads
\begin{equation}
\begin{aligned}
 V = \ &\tfrac{\alpha'e^{\phi}}{{\mathcal V}} (X^n \cdot X^m) \int \Bigl({\mathcal F}_+ ^n + \sum \limits_i q^n _i 
\overline{\Phi}^{\overline{\bf R}_i} _{k_i} (\U_{k_i l_i}) \Phi^{{\bf R}_i} _{l_i} \Bigr) 
\wedge \star \Bigl({\mathcal F}_+ ^m + \sum \limits_i q^m _i 
\overline{\Phi}^{\overline{\bf R}_i} _{k_i} (\U_{k_i l_i}) \Phi^{{\bf R}_i} _{l_i} \Bigr) 
 \\
&+ \tfrac{\alpha'e^{\phi}}{{\mathcal V}} \sum \limits_a \int \Bigl( \sum
   \limits_i \overline{\Phi}^{\overline{\bf R}_i} _{k_i} (\U_{k_i
     l_i}) (\tau^a \Phi)^{{\bf R}_i} _{l_i} \Bigr) \wedge \star \Bigl( \sum
   \limits_i \overline{\Phi}^{\overline{\bf R}_i} _{k_i} (\U_{k_i
     l_i}) (\tau^a \Phi)^{{\bf R}_i} _{l_i} \Bigr) \ ,
\end{aligned}
\end{equation}
where $q_i ^n$ is the charge of the field $\Phi^{{\bf R}_i}$ under the group $U(1)_n$.

Recalling the general argument in section \ref{dev}, the rigid fluxes of the line bundle background stabilize some of the $K3$ moduli. 
The Fayet-Iliopoulos term ${\mathcal F}_+$ in \eqref{FI} is generated by those $K3$ metric deformations that violate the ASD condition of the Yang-Mills background. 
Hence, their mass is lifted to a nonzero value. 
Since we have an Abelian gauge flux in the case of line bundles, i.e.\ ${\mathcal F} \in H^2(K3, \mathbb{Z})$, 
we get an intuitive picture of the moduli stabilization in terms 
of the 3-plane $\Sigma \in H^2(K3, \mathbb{R})$, introduced in section \ref{K3}. 
The ASD condition \eqref{hym2} can be written as 
\begin{equation}\label{perp}
 {\mathcal F} \perp \Sigma \ ,
\end{equation}
where orthogonality is defined with respect to the intersection matrix $\rho$. 
Hence, massless deformations of the $K3$ metric are given by all motions of $\Sigma$, preserving \eqref{perp}. 
For $N$ line bundles the massless metric deformations are constrained to the subspace orthogonal to the flux vectors $\{{m}^1,..., {m}^N\}$.
If all $N$ flux vector are linearly independent, the remaining moduli space is described by the Grassmannian manifold
\begin{equation}\label{redmod}
 \tilde{{\mathcal M}}_{K3} = \frac{O(3,19-N)}{O(3) \times O(19-N)} \times \mathbb{R}^+,
\end{equation}
so there are $3N$ moduli stabilized and $\mbox{dim} \ \tilde{{\mathcal
    M}}_{K3} = 58-3N$. For $E_8 \times E_8$ we have $N_{max} = 16$,
which stabilizes all but 10 moduli and leaves $U(1)^{16}$
unbroken. For a GUT group to survive in 6D
a larger number of moduli has to stay unfixed. 

Finally, let us mention that there exists also a moduli space for the
charged scalars which consists of all $D$-flat directions $C^{\bf R}
_{k_i}, D^{\bf R} _{k_i} \neq 0$, satisfying $D^a=D^{\bf 1}=0$.
The corresponding Higgs branch has a smaller  gauge group with less
massless hypermultiplets~\cite{Kachru:1996pc}.

\subsection{St\"uckelberg mechanism and massive U(1)s}

We close this paper by analyzing the effect of the affinely gauged
scalars $b^I$ (cf.~\eqref{db2}).
Let us first focus on one line bundle for simplicity. In this case 
the $U(1)$ gauge symmetry acts according to 
\begin{equation}
\begin{aligned}
 V^{\bf 1} &\longrightarrow V^{\bf 1} + d\chi \ , \qquad
 b^I \longrightarrow b^I + \alpha' m^I \chi\ .
\end{aligned}
\end{equation}
This implies that
one combination of $b^I$ can be gauged to zero with $V^{\bf 1}$
becoming massive which is known as the St\"uckelberg mechanism.\footnote{In 6D this effect is independent of possible Abelian anomalies \cite{Honecker:2006dt}.} 
The mass term (in the Einstein frame) is found from \eqref{db2} to be 
\begin{equation}
\tfrac{\alpha'^2}{6 \mathcal V}  \lVert \flux \rVert^2 V^{\bf 1} \wedge * V^{\bf 1} \int \mbox{tr} ({\mathcal F} \wedge \star {\mathcal F}) 
= - \tfrac{\alpha'^2}{6 \mathcal V}  \lVert \flux \rVert^4 V^{\bf 1} \wedge * V^{\bf 1} \rho_{IJ} m^I m^J \ ,
\end{equation}
where we used the ASD condition $\star {\mathcal F} = -{\mathcal F}$.
To identify the physical mass we need to absorb a factor
$\sqrt{\alpha'} \lVert V \rVert$ into $V^{\bf 1}$ in order to get a
canonical kinetic term as can be seen from \eqref{Kinetic6Dterms}. 
Using the tadpole condition \eqref{tadpole} the physical mass reads
\begin{equation}
 m = 4 \sqrt{\tfrac{\alpha'}{\mathcal V}} \ .
\end{equation}
Note that the physical mass only depends on the $K3$ volume.

If there are $N$ line bundles with flux parameters ${m}^{In} = (m^{I1},..., m^{IN})$, the $b^I$ are coupled to all of them and generically all 
``fluxed" $U(1)$'s become massive. However, if some flux vectors are linearly dependent, $\mbox{dim span}\{m^1,..., m^N\} = K < N$, the rank of the mass matrix is reduced and there remain $N-K$ massless $U(1)$'s in the spectrum. 
Let us show which combination of $b_I$-scalars is eaten by which combination of $U(1)$'s. In an integral basis of $H^2(K3, \mathbb{Z})$ we define 
$q^{In} = \lVert V^n \rVert m^{In} \in \mathbb{Z}$ and look for the orthogonalization
\begin{equation}
{\mathcal L}_6 \sim g_{IJ} (db^I - q^{In} V_n ^{\bf 1})^2 = \tilde{g}_{IJ} (d\tilde{b}^I - \lambda^{In} \tilde{V}_n ^{\bf 1} )^2 \ .
\end{equation}
For $K$ linear independent flux vectors the $22 \times N$ matrix $q^{In}$ has rank $K$ and hence can be be brought to the following form (e.g.\ $N=3, K=2$)
\begin{equation}
 q^{In} \mapsto O^I _{\ J} q^{Jm} U^n _{\ m} = \lambda^{In} = \begin{pmatrix} \lambda^1 & 0 & 0 & \ldots \\ 0 & \lambda_2 & 0 & \ldots \\ 0 & 0 & 0 & \ldots \end{pmatrix} \ ,
\end{equation}
where $O \in O(22)$ and $U \in O(N)$.
This determines the preferred basis 
\begin{equation}
 \tilde{V}_n ^{\bf 1} = U_n ^{\ p} V_p ^{\bf 1} \ , \qquad \tilde{b}^I = O^I _{\ J} b^J \ ,
\end{equation}
in which the first $K$ $\tilde{b}$ scalars are the Goldstone bosons of the first $K$ gauge potentials. 
More precisely, one goes to a basis of $H^2(K3, \mathbb{Z})$ where the flux hyperplane $\mbox{span}(m^1,\ldots, m^n)$
is spanned by the first $K$ harmonic 2-forms $\tilde{\eta}_1, \ldots \tilde{\eta}_K$. 
The special form of $\lambda^{In}$ however does not tell us if this basis is orthogonal with respect 
to the intersection matrix \eqref{insect}. 
Since we have $\star {\mathcal F}^n = - {\mathcal F}^n$ for each gauge flux, the mass terms read
\begin{equation}
 \tfrac{\alpha'}{6 \mathcal V} {V}_n ^{\bf 1} \wedge * {V}_m ^{\bf 1} \int {\mathcal F}^n \wedge \star {\mathcal F}^m
 = -\tfrac{\alpha'}{6 \mathcal V} \tilde{V}_n ^{\bf 1} \wedge * \tilde{V}_m ^{\bf 1} \tilde{\rho}_{IJ} \lambda^{In} \lambda^{Jm} \ ,
\end{equation}
where $\tilde{\rho}_{IJ} = O_I ^{\ K} O_J ^{\ L} \rho_{KL}$.
In general $\tilde{\rho}_{IJ}$ will not be diagonal and hence the mass term will not be diagonal in $n,m$. 
Therefore, the mass eigenbasis is generically different from the ``Goldstone eigenbasis''. 
Note that again the mass matrix only depends on the volume modulus and that the trace of the (squared) mass matrix is fixed by the tadpole condition
\begin{equation}
 \mbox{tr}(M^2) = \sum\limits_n (-\tfrac13 \tfrac{\alpha'}{\mathcal V} \tilde{\rho}_{IJ} \lambda^{In} \lambda^{Jn}) = 16\, \tfrac{\alpha'}{\mathcal V} \ .
\end{equation}

\section{Conclusion}

In this paper we derived the six-dimensional
low energy effective action of
the heterotic string compactified on
$K3$.
Consistency requires a nontrivial gauge bundle
 on $K3$ and for concreteness we chose to  
consider first the standard embedding and second a flux background with
$U(1)$ line bundles.
In both cases we performed a Kaluza-Klein reduction starting from 
the ten-dimensional action.
Specifically we focused on the gauge sector where charged and neutral 
scalars (bundle moduli) arise as massless deformations of the internal gauge
bundle.
We carefully performed a KK-reduction and computed the
sigma-model metric and the scalar potential of the six-dimensional action 
as a functions of the geometrical
$K3$ moduli and the axionic scalars arising from the NS $B$-field.
For the scalar potential we showed the
consistency with the generic 6D, ${\mathcal N} = 1$ supergravity
in that it arises solely from a $D$-term.
The sigma-model metric is constrained to be a quaternionic-K\"ahler
metric which, however, we could only show in an appropriate orbifold
limit.
The proof that the full metric computed in this paper is indeed 
quaternionic-K\"ahler is left for a future project.

The line bundle backgrounds are realized by Abelian Yang-Mills fluxes
on $K3$.
They affect the 6D theories in that the scalars arising from the $B$-field
become affinely gauged under the unbroken $U(1)$'s.
This in turn gives a mass to the $U(1)$ gauge fields via a
St\"uckelberg mechanism.
For several line bundles which are linearly dependent in $H^2(K3,
\mathbb{Z})$, massless $U(1)$ gauge fields remain in the 6D theory.
At the same time the fluxes stabilize those $K3$ moduli
which violate the anti-selfduality of the Yang-Mills field
strength.
In the effective potential this is realized as a Fayet-Iliopoulos term
proportional to the flux vector.
Together, one line bundle eliminates four scalars (one $B$ scalar and
three $K3$ moduli) from the effective theory, which are absorbed into a
massive vector multiplet.

Recently \cite{Bonetti:2011mw} derived the 6D effective action of F-theory
compactified on a Calabi-Yau three-fold $X$. When $X$ is a $K3$ fibration,
this background is dual to the heterotic theory compactified on $K3$ studied
in this paper. It would be interesting to compare the two effective actions. On the
F-theory side one may use our results to get information on the
couplings of the charged matter
(in \cite{Bonetti:2011mw} the action was derived on a generic point in the Coulomb branch, where these
fields are massive, but eventually one has to go away from this branch in the F-theory
limit). On the heterotic side one may use the results of \cite{Bonetti:2011mw}
to understand the couplings of non-perturbative
tensors (that in F-theory appear at perturbative level).

\vskip 1cm

\subsection*{Acknowledgments}

This work was supported by the German Science Foundation (DFG) within
the Collaborative Research Center (SFB) 676. We have greatly benefited
from conversations and correspondence with 
A.P.~Braun, W.~Buchm\"uller,  A.~Collinucci, V.~Cort\'es, T.~Grimm, 
S.~Groot Nibbelink,
G.~Honecker,
J.~ Schmidt and C.~Scrucca.

\vskip 1cm

\begin{appendix}

\section{Details of the Kaluza-Klein reduction}\label{def}
\subsection{Deformations of gauge connections}\label{def1}
In this appendix 
we give a detailed derivation of the Kaluza-Klein reduction of the gauge potential, from which all bosonic matter fields descend. The low energy spectrum is determined by the gauge 
background consisting of a nontrivial holomorphic $H$-bundle over $K3$ and a flat $G$-bundle over $M^{1,5}$
\begin{equation}
 E_8 \times E_8 \longrightarrow G \times \langle H \rangle \ ,
\end{equation}
where
$G$ is the maximal commutant of $H$. The $H$-bundle satisfies the Bianchi identity \eqref{BI2} and its nonzero field strength 
${\mathcal F}$ satisfies the hermitean Yang-Mills equations (HYM)
\begin{equation}\label{hym1a}
 {\mathcal F} \in H^{1,1}(K3, \mathfrak{h}), \ \ \ \ {\mathcal F} \wedge J = 0\ .
\end{equation}
Here we write $\mathfrak h$ for the adjoint $H$-bundle. \eqref{hym1a}
is equivalent to the anti-selfduality (ASD) of the field strength, 
$\star {\mathcal F} = - {\mathcal F}$ \cite{Atiyah:1978wi, Friedman:1998tb}.
We denote the background connection, valued in ${\mathfrak h}$, 
as ${\mathcal A}$
and its deformations give rise to massless 6D 
fields.\footnote{Since we insist on six-dimensional Lorentz
  invariance we do not include the possibility of
a background value for the 6D gauge field.}
%
These deformations are grouped into multiplets according to the decomposition
\begin{equation}\label{dec}
 {\bf 496} \rightarrow \bigoplus_i \ ({\bf R}_i,{\bf S}_i) \oplus ({\mathfrak g}, {\bf 1}) \oplus ({\bf 1}, {\mathfrak h})\ ,
\end{equation}
where ${\mathfrak g}$ and ${\mathfrak h}$ denote the adjoint representations of $G$ and $H$, respectively. The ${\bf 1}$ is the trivial representation and 
$({\bf R}_i,{\bf S}_i)$ are group specific representations.
It is known from supersymmetry that massless 6D hypermultiplets in
representations ${\bf R}_i$ occur with multiplicities given by the
chiral index \cite{Green:1984bx}
\begin{equation}\label{index}
 \chi(E_{{\bf S}_i}) = h^{0,0}(K3, E_{{\bf S}_i}) - h^{0,1}(K3, E_{{\bf S}_i}) + h^{0,2} (K3, E_{{\bf S}_i}) \ ,
\end{equation}
where $E_{{\bf S}_i}$ denotes the vector bundle associated with ${\bf S}_i$.\footnote{$\chi$ is called chiral index due to the equivalent definition $\chi(E) = n^+ _E - n^- _E$, 
where $n_E ^{\pm}$ count the chiral zero modes of the Dirac operator.
On $K3$ one has $\chi(E) = \chi(E^*)$, so complex conjugate representations always
occur with equal multiplicities. Due to the definite chiralities in
the vector- and hypermultiplets, $\chi(E)$ counts the difference of
them.} 
In fact, $h^{0,0}(K3, E)$ and $h^{0,2} (K3, E)$ vanish for a HYM background.
This can be seen as follows: $H^{0,0}(K3, E)$ is the space of global
sections of $E$, which are closed with respect to the covariant
Dolbeault 
operator $\bar{\partial}_{\mathcal A}$  on $K3$.
But for sections of a HYM-bundle we have the identity\footnote{A proof can 
be found, for example, in appendix E of \cite{Beasley:2008dc}.} 
\begin{equation}\label{id}
d_{\mathcal A} ^*
d_{\mathcal A} = 2 \bar{\partial}_{\mathcal A} ^*
\bar{\partial}_{\mathcal A}\ ,
\end{equation}
where $d_{\mathcal A}= \partial_{\mathcal A} +
\bar{\partial}_{\mathcal A}$.
Therefore any such section is also covariantly
constant. 
When $E$ is nontrivial and irreducible, no constant sections
exist. The vanishing of $H^{0,2} (K3, E)$ then follows by Serre
duality \cite{Hubsch:1992nu}.

For the Kaluza-Klein reduction of the bosonic action it is not enough to know this multiplicity. 
One has to know which internal differential equation the zero modes satisfy. 
Therefore we analyze the deformations of the gauge connection without referring to supersymmetry.
Starting from the 10D Yang-Mills Lagrangian 
\begin{equation}\label{YM}
 {\mathcal L}^{YM} \sim \langle F, F \rangle = \mbox{tr} (F \wedge * F) \ ,
\end{equation}
we parametrize the deformations by $A=  {\mathcal A} + \dA $ with $a \in \Lambda^1 ({\mathfrak e}_8)$. 
For simplicity we assume that the background $H$-bundle is inside one $E_8$
and consider only deformations inside this $E_8$. 
We restrict $a$ to be compatible with the metric on the adjoint $E_8$ bundle.\footnote{This amounts to the condition that the deformed connection $A = {\mathcal A} + a$ satisfies 
$d(h(\psi_1, \psi_2)) = h(d_A \psi_1, \psi_2) + h(\psi_1, d_A \psi_2)$, where $h$ is the adjoint metric, i.e.\ locally the Killing form of the Lie algebra, and $\psi_1, \psi_2$ are sections of the adjoint bundle.}
The field strength deforms as
\be
F = {\mathcal F} + \dF \ ,\qquad
\dF = d_{\mathcal A} \dA + \tfrac12 [\dA,\dA]\ .
\ee
As in the main text we decompose $\dA=\dA_1+\dA_{\bar 1}$
into 1-forms on $M^{1,5}$ and on $K3$. They deform the flat $G$- and the curved $H$-connection, respectively. Their 6D effective mass terms 
are given by 
\begin{equation}\label{mass1}
 {\mathcal L}_6 ^{mass}[\dA_1] \sim \int \limits_{K3} \mbox{tr} (d_{\mathcal A} \dA_1 \wedge \star d_{\mathcal A} \dA_1) \ ,
\end{equation}
\begin{equation}\label{mass2}
 {\mathcal L}_6 ^{mass} [\dA_{\bar 1}] \sim \int \limits_{K3} \mbox{tr} (d_{\mathcal A} \dA_{\bar 1} \wedge \star d_{\mathcal A} \dA_{\bar 1}) + 
\int \limits_{K3} \mbox{tr} (\dA_{\bar 1} \wedge \star [{\mathcal F}, \dA_{\bar 1}]) \ .
\end{equation}
From \eqref{mass1} it follows that massless 6D vectors $V_i$ arise from deformations with $d_{\mathcal A} \dA_1 = 0$. Therefore the Kaluza-Klein expansion reads
\begin{equation}
 \dA_1 = V \cdot \psi \ , \qquad d_{\mathcal A} \psi = 0 \ ,
\end{equation}
with internal covariantly constant functions (sections) $\psi$. Since there exist no globally constant sections on nontrivial vector bundles,
massless 6D vectors can only occur from the term $({\mathfrak g}, {\bf
  1})$ in \eqref{dec}. From  the identity \eqref{id}  
(on sections) it follows that $\mbox{ker}(d_{\mathcal A}) = 
\mbox{ker}({\bar \partial}_{\mathcal A})$. Hence the multiplicity is given by Dolbeault cohomology 
\begin{equation}\label{vec}
h^{0,0}(K3, E_{\bf 1}) = h^{0,0}(K3) = 1 \ .
\end{equation}

The mass operator for 6D scalars is identified from \eqref{mass2} as 
\begin{equation}\label{mass3}
 \Delta_{YM} \dA_{\bar 1} := d^* _{\mathcal A} d_{\mathcal A} \dA_{\bar 1} + \star [{\mathcal F}_{\mathcal A}, \dA_{\bar 1}]  \ .
\end{equation}
Since this is not a proper Laplacian, the connection to Dolbeault cohomology is obscure at first sight. 
We now show that 1-form zero modes of $\Delta_{YM}$ are in one-to-one correspondence with 
zero modes of $\Delta_{{\bar \partial}_{\mathcal A}} := {\bar \partial}_{\mathcal A} ^* {\bar \partial}_{\mathcal A} + {\bar \partial}_{\mathcal A} {\bar \partial}_{\mathcal A} ^*$. 
Using the K\"ahler identities ${\bar \partial}_A ^* = i[\partial_A, J
\cdot]$ and $\partial_A ^* = -i[{\bar \partial}_A, J \cdot]$
 \cite{Huy}, we find
the following operator identity on 1-forms
\begin{equation}\label{id1}
 d_{\mathcal A} ^* d_{\mathcal A} \dA_{\bar 1} = 2 \Delta_{{\bar \partial}_{\mathcal A}} \dA_{\bar 1} - d_{\mathcal A} d_{\mathcal A} ^* \dA_{\bar 1} + i J \cdot [{\mathcal F}, \dA_{\bar 1}] \ .
\end{equation}
Here $J \cdot$ is the contraction with the K\"ahler form. (There is an equivalent identity with $\Delta_{{\partial}_{\mathcal A}}$ instead of $\Delta_{{\bar \partial}_{\mathcal A}}$.)
We prove \eqref{id1} at the end of this section.
The second term on the r.h.s.\ vanishes in the Lorenz gauge $d_{\mathcal A} ^* \dA_{\bar 1} = 0$.
Moreover, on a complex K\"ahler surface with a HYM-bundle (i.e.\
anti-selfdual field strength) one can show that
\begin{equation}\label{id2}
 \star [{\mathcal F}, \dA_{\bar 1}] = -i J \cdot [{\mathcal F}, \dA_{\bar 1}] \ .
\end{equation}
Inserting \eqref{id1} and \eqref{id2} into the mass operator \eqref{mass3}, we are left with the (gauge fixed) identity on 1-forms
\begin{equation}\label{YM2}
 \Delta_{YM} = 2 \Delta_{{\bar \partial}_{\mathcal A}} = 2 \Delta_{{\partial}_{\mathcal A}} \ .
\end{equation}

Since on holomorphic bundles the Dolbeault operator satisfies ${\bar
  \partial}_{\mathcal A} ^2 = 0$, the harmonic 1-forms of $\Delta_{{\bar \partial}_{\mathcal A}}$ are unique representatives 
of $H^{0,1}(K3, E)$. From \eqref{YM2} it also follows that the massless modes are zero modes of $d_{\mathcal A}$. This is obvious from 
\eqref{mass3} as a sufficient condition, but here we have shown that it is also necessary. Another way of seeing this is the following: Whereas the first term in \eqref{mass3} is a positive, 
symmetric operator, the second is in fact antisymmetric with respect to the YM-scalar product on $K3$
\begin{equation}
\langle \dA_{\bar 1}, \star [{\mathcal F}, \dA_{\bar 1}] \rangle = - \langle \star [{\mathcal F}, \dA_{\bar 1}], \dA_{\bar 1} \rangle \ .
\end{equation}
Hence, the two terms correspond to real and imaginary part of the squared mass eigenvalues and have to vanish separately. 
%
Hence, we derived the supersymmetric result from pure bosonic Yang-Mills deformation theory. 

Returning to the different terms in \eqref{dec}, no 6D scalars in the adjoint representation $\mathfrak g$ can occur, because $H^{0,1}(K3, E_{\bf 1}) = H^{0,1}(K3) = 0$. 
Generically one gets scalars from representations $({\bf R}, {\bf S})$ with some multiplicity $h^{0,1}(K3, E_{{\bf S}})$. Here two cases can arise: First, if $({\bf R}, {\bf S})$ is a real
representation and ${\bf R}$ is pseudoreal and we are left with ${\bf R}$-half-hypermultiplets in 6D.
To have complex fields in 6D one decomposes the deformation as $a_{\bar 1} = a^{0,1} + a^{1,0}$, using a complex structure on $K3$.
Since $a$ is restricted to preserve the hermitean structure of the ${\mathfrak e}_8$ bundle, the two terms satisfy \cite{Huy}
\begin{equation}\label{antih}
 (a^{1,0})^{\dagger} = - a^{0,1} \ .
\end{equation}
Hence, the Kaluza-Klein expansion reads
\begin{equation}
 a_{\bar 1} = C_k ^{\bf R} \omega_k + \overline{C}_k ^{\overline{\bf R}} \overline{\omega}_k \ .
\end{equation}
Second, if there are complex representations occuring in conjugated pairs, $({\bf R}, {\bf S}) \oplus (\overline{\bf R}, \overline{\bf S})$, 
two sets of independent 6D scalars arise 
\begin{equation}\label{agenau}
 \dA_{\bar 1} = C^{{\bf R}} _k \omega_k + \overline{C}^{\overline{\bf R}} _k \overline{\omega}_k + D^{{\bf R}} _k \varpi_k + \overline{D}^{\overline{\bf R}} _k \overline{\varpi}_k \ .
\end{equation}
The zero modes of both cases are given by
\begin{equation}
\begin{aligned}
 \omega_k \in H^{0,1}(K3, E_{{\bf S}}) \ ,& \qquad \overline{\omega}_k \in H^{1,0}(K3, E_{\bar{\bf S}}) \ , \\
 \varpi_k \in H^{1,0}(K3, E_{{\bf S}}) \ ,& \qquad \overline{\varpi}_k \in H^{0,1}(K3, E_{\bar{\bf S}}) \ .
\end{aligned}
\end{equation}
Here $E_{\bar{\bf S}} = (E_{{\bf S}})^*$ is the dual vector bundle. On $K3$ all multiplicities are the same due to Serre duality
\begin{equation}
 \overline{H^{0,1} (K3, E_{\bf S})} \cong H^{0,1}(K3, E_{\bar{\bf S}}) 
\end{equation}
and can be computed via the chiral index \eqref{index}.\footnote{On a Calabi Yau 3-fold the $C^{\bf R}$ and $\overline{D}^{\overline{\bf R}}$ 
occur with different multiplicities, yielding the 4D chiral spectrum.}  
Thus, in 6D one has hypermultiplets with scalar components $\Phi_k ^{{\bf R} \oplus {\overline{\bf R}}} = (C_k ^{\bf R}, {\overline D}_k ^{\overline{\bf R}})$.

Let us now show that 
the 6D singlet scalars coming from the term $({\bf 1}, {\mathfrak h})$
in \eqref{dec} are  special in that they are not only massless but
exact flat directions of the potential. They are termed bundle moduli.
Applying the previous analysis it follows that there exist massless
deformations with multiplicity $h^{0,1}(K3,{\mathfrak h})$. In fact,
any such deformation preserves \eqref{hym1a} and hence the ASD
condition of the background $\mathcal F$. 
It is known that the moduli space of ASD connections modulo gauge
transformations  is equivalent to the moduli space of holomorphic structures (see for example \cite{Friedman:1998tb}).
A holomorphic structure is defined by a Dolbeault operator satisfying
 $\bar{\partial}_{\mathcal A} ^2 = {\mathcal F}^{0,2} = 0$.
A deformation $A = {\mathcal A} + a$, with $a \in \Lambda^1(K3, {\mathfrak h})$ defines another holomorphic structure if ${\mathcal F}_A ^{0,2} = 0$, i.e.\
\begin{equation}\label{eq}
\bar{\partial}_{\mathcal A} a^{0,1} + \tfrac12 [a^{0,1}, a^{0,1}] = 0 \ .
\end{equation}
Infinitesimally this yields $a^{0,1} \in \mbox{ker}
(\bar{\partial}_{\mathcal A})$.
However $a \in \mbox{ker}(\bar{\partial}_{\mathcal A})$ contains directions which lead to gauge-equivalent holomorphic structures which have to be modded out.
Their Dolbeault operators are related by conjugation in $H$
\begin{equation}
\bar{\partial}_{\mathcal A} ^{h} = h^{-1} \bar{\partial}_{\mathcal A}
h \approx \bar{\partial}_{\mathcal A} + \bar{\partial}_{\mathcal A}
\delta h \ ,
\end{equation}
where $h\in \Lambda^0(K3, H)$ and $h \approx {\bf 1} + \delta h, \ 
\delta h \in \Lambda^0(K3, {\mathfrak h})$.
Modding out the term $\bar{\partial}_{\mathcal A} \delta h \in \mbox{Im}(\bar{\partial}_{\mathcal A})$, infinitesimal deformations of the holomorphic structure are given by $a^{0,1} \in H^{0,1}(K3, {\mathfrak h})$, in agreement with the result from the mass operator. But since the effective scalar potential from the background takes the form
\begin{equation}
V_6 \sim - \int \mbox{tr} ({\mathcal F}_+ \wedge \star {\mathcal F}_+) \ , 
\end{equation}
(see \eqref{shift2}) all deformations preserving the ASD condition are 
moduli, i.e.\ flat directions of the scalar potential. 
Finally, the Kaluza-Klein expansion of the $({\bf 1}, {\mathfrak h})$-scalars reads
\begin{equation}\label{gaugec}
a_{\bar 1} = \xi_k \alpha_k + \overline{\xi}_k  \overline{\alpha}_k \ , \qquad \alpha_k \in H^{0,1}  (K3, {\mathfrak h}) \ .
\end{equation}
The complex 6D scalars $\xi_k$ are called bundle moduli. 
In the following sections the above results are applied to the standard embedding and the line bundle background.

We finally give a proof of the formula \eqref{id1} for $a \in \Lambda^1 (K3,E)$:
\begin{equation}
 \begin{aligned}\label{int}
  d^* _{\mathcal A} d_{\mathcal A} a &= (\bar{\partial}^* _{\mathcal A} \bar{\partial}_{\mathcal A} + \partial^* _{\mathcal A} \partial_{\mathcal A})a + (\bar{\partial}^* _{\mathcal A} \partial_{\mathcal A} + \partial^* _{\mathcal A} \bar{\partial}_{\mathcal A})a \\
 \end{aligned}
\end{equation}
The first term can be written as
\begin{equation}\label{I}
\begin{aligned}
(\bar{\partial}^* _{\mathcal A} \bar{\partial}_{\mathcal A} + \partial^* _{\mathcal A} \partial_{\mathcal A})a &=
i([\partial_{\mathcal A}, J \cdot] \bar{\partial}_{\mathcal A} - [\bar{\partial}_{\mathcal A}, J \cdot]
\partial_{\mathcal A})a \\
&= i(\partial_{\mathcal A} J \cdot \bar{\partial}_{\mathcal A}  - \bar{\partial}_{\mathcal A} J \cdot \partial_{\mathcal A}) a 
 - i J \cdot (\partial_{\mathcal A} \bar{\partial}_{\mathcal A} a - \bar{\partial}_{\mathcal A} \partial_{\mathcal A} a) \\
 &= (\partial_{\mathcal A} \partial^* _{\mathcal A} + \bar{\partial}_{\mathcal A} \bar{\partial}^* _{\mathcal A})a + i J \cdot [{\mathcal F}, a] - 2i J \cdot (\partial_{\mathcal A} \bar{\partial}_{\mathcal A} a) \ .
\end{aligned}
\end{equation}
Here we used the K\"ahler identities $\partial^* _{\mathcal A} =
-i[\bar{\partial}_{\mathcal A}, J \cdot]$, $\bar{\partial}^* _{\mathcal A} = i[\partial_{\mathcal A}, J
\cdot]$, $J\cdot \partial_{\mathcal A} a = [J\cdot, \partial_{\mathcal A}] a$ since $J\cdot a = 0$, and we identified ${\mathcal F} = \partial_{\mathcal A} \bar{\partial}_{\mathcal A} + \bar{\partial}_{\mathcal A} \partial_{\mathcal A}$.
We now write the last term in \eqref{I} as
\begin{equation}
\begin{aligned}
 2i J \cdot (\partial_{\mathcal A} \bar{\partial}_{\mathcal A} a) &= 2i([J\cdot, \partial_{\mathcal A}] + \partial_{\mathcal A} J \cdot) \bar{\partial}_{\mathcal A} a \\
 &= -2 \bar{\partial}^* _{\mathcal A} \bar{\partial}_{\mathcal A} a + 2i \partial_{\mathcal A} [J \cdot, \bar{\partial}_{\mathcal A}] a \\
 &= -2 \bar{\partial}^* _{\mathcal A} \bar{\partial}_{\mathcal A} a + 2 \partial_{\mathcal A} \partial^* _{\mathcal A} a \ .
\end{aligned}
\end{equation}
With this we get 
\begin{equation}
 (\bar{\partial}^* _{\mathcal A} \bar{\partial}_{\mathcal A} + \partial^* _{\mathcal A} \partial_{\mathcal A})a = (\bar{\partial}_{\mathcal A} \bar{\partial}^* _{\mathcal A} - \partial_{\mathcal A} \partial^* _{\mathcal A})a + iJ \cdot [{\mathcal F}, a] + 2 \bar{\partial}^* _{\mathcal A} \bar{\partial}_{\mathcal A} a \ .
\end{equation}
Now we consider the second term in \eqref{int}
\begin{equation}
(\bar{\partial}^* _{\mathcal A} \partial_{\mathcal A} + \partial^* _{\mathcal A} \bar{\partial}_{\mathcal A})a= ( \partial_{\mathcal A} \bar{\partial}^* _{\mathcal A} + \bar{\partial}_{\mathcal A} \partial^* _{\mathcal A} ) a = d_{\mathcal A} d^* _{\mathcal A} a - (\partial_{\mathcal A} \partial^* _{\mathcal A} + \bar{\partial}_{\mathcal A} \bar{\partial}^* _{\mathcal A})a \ ,
\end{equation}
where  we used $\{\partial_{\mathcal A}, \bar{\partial}^* _{\mathcal A}\} = 0$ (which follows from the K\"ahler identities). 
Together we end up with the claimed result \eqref{id1}
\begin{equation}
 d_{\mathcal A} ^* d_{\mathcal A} a = - d_{\mathcal A} d_{\mathcal A} ^* a + 2(\bar{\partial}^* _{\mathcal A} \bar{\partial}_{\mathcal A} + \bar{\partial}_{\mathcal A} \bar{\partial}^* _{\mathcal A})a + i J \cdot [{\mathcal F}, a] \ .
\end{equation}

\subsection{Zero modes in the standard embedding}\label{zeroSE}
For the standard embedding the nontrivial $SU(2)$ bundle is inside one $E_8$ factor, yielding the breaking
\begin{equation}
 E_8 \longrightarrow E_7 \times \langle SU(2) \rangle \ .
\end{equation}
Focusing on this $E_8$ factor we have the decomposition
\begin{equation}
 {\bf 248} \rightarrow ({\bf 56},{\bf 2}) \oplus ({\bf 133},{\bf 1}) \oplus ({\bf 1},{\bf 3}) \ .
\end{equation}
The vector bundles $E$ corresponding to the right entries are identified as
$E_{\bf 2} = {\mathcal T}_{K3}$, which is the holomorphic tangent bundle, $E_{\bf 3}=\mathfrak{su}(2)=\mbox{End}\ {\mathcal T}_{K3}$, which is the adjoint bundle and 
$E_{\bf 1} = \mathcal{O}$, which is the trivial bundle over $K3$. Since $(\bf 56, \bf 2)$ is a real representation, its massless Kaluza-Klein components are given by
\begin{equation}
 \dA_{\bar 1} ^{({\bf 56}, {\bf 2})} = C^{\bf 56} _j \omega_j + \overline{C}^{\overline{\bf 56}}_j
 \overline{\omega}_j \ ,\qquad j=1,\ldots,20\ .
\end{equation}
Here the zero modes are 
\begin{equation}
\begin{aligned}
 \omega_j \in H^{0,1} ({\mathcal T}_{K3}) \cong H^{1,1} (K3) \ , \\ 
 \overline{\omega}_j \in {H^{1,0} (\overline{\mathcal T_{K3}})} \cong H^{1,1} (K3) \ .
\end{aligned}
\end{equation}
From the Hodge diamond \eqref{dia} we see that the multiplicity is 20. 
We realize the isomorphy to $H^{1,1} (K3)$ with the holomorphic 2-form $\Omega$ and a particular prefactor, 
i.e.\ in components (no summation over $j$ implied) \footnote{There exists an alternative isomorphism, $\omega_{\bar \alpha} ^{\beta} \propto g^{\beta \bar{\gamma}} 
(t_{(\bar{\alpha} \bar{\gamma})} + t_{[\bar{\alpha} \bar{\gamma}]}) = g^{\beta \bar{\gamma}} (\overline{\Omega}^{\delta} _{\ (\bar{\alpha}} \omega_{\bar{\gamma}) \delta} + 
\overline{\Omega}_{\bar{\alpha} \bar{\gamma}})$, which maps $H^{0,1}({\mathcal T}_{K3})$ to the anti-holomorphic 2-form $\overline{\Omega}$ plus all $(1,1)$-forms except the K\"ahler form. 
We always use the simpler one \eqref{zero1}.}
\begin{equation}\label{zero1}
\begin{aligned}
 (\omega_j)_{\bar \alpha} ^{ \ \beta} &= \tfrac{\gamma_j}{|| \Omega ||^2} \overline{\Omega}^{\alpha \beta} (\eta_j)_{\alpha \bar{\alpha}} \ , \\
 (\overline{\omega}_j)_{\alpha} ^{\ \bar{\beta}} &= \tfrac{\gamma_j}{|| \Omega ||^2} \Omega^{\bar{\alpha} \bar{\beta}} (\eta_j)_{\alpha \bar{\alpha}} \ ,
\end{aligned}
\end{equation}
where $\eta_j$ are the harmonic $(1,1)$ forms on $K3$ and $\gamma_j$ is the real function
\begin{equation}\label{gi}
 \gamma_j = \frac{{\mathcal V}^{\frac{1}{4}}}{\left(\int J \wedge \eta_j\right)^{\frac12}} \ .
\end{equation}
This function is motivated by matching with the orbifold limit of the standard embedding which we discuss in appendix \ref{limit}. 
In fact, the zero modes of the charged scalars depend on the complex structure of $K3$ by the very definition of ${\mathcal T}_{K3}$.
For a fixed complex structure the prefactor $\gamma_j$ depends on the remaining K\"ahler moduli in such a way that the full zero mode is independent of them.

The term $({\bf 133}, {\bf 1})$ gives rise to one 6D vector $V^{\bf 133}$, as
stated in \eqref{vec}. 
The term $({\bf 1}, {\bf 3})$ corresponds to the bundle moduli as specified in \eqref{gaugec}
\begin{equation}
 \dA_{\bar 1} ^{({\bf 1}, {\bf 3})} = \xi_k \alpha_k + \overline{\xi}_k \overline\alpha_k  \ , \qquad \alpha_k \in H^{0,1} (\mbox{End} \ {\mathcal T}_{K3}) \ .
\end{equation}
 The multiplicity cannot be related to the Hodge numbers but can be computed with the chiral index \eqref{index}.
Here $h^{0,0} (\mbox{End} \ {\mathcal T}_{K3}) = 0$, since a covariantly constant section $g \in \Gamma(K3, \mbox{End} \ {\mathcal T}_{K3})$ must take values in the centralizer of the 
holonomy group, which is empty for $\mathfrak{hol}(K3) =
\mathfrak{su}(2)$
\cite{Atiyah:1978wi}.
Thus, one obtains
\begin{equation}
\chi (\mbox{End} \ {\mathcal T}_{K3}) = -h^{0,1} (\mbox{End} \ {\mathcal T}_{K3})\ .
\end{equation}
$\chi$ can be computed via the Hirzebruch-Riemann-Roch theorem\footnote{See for example chapter 5.1 of \cite{Huy}.}  which states
\begin{equation}
 \chi(E_{\bf S}) = \int \limits_{K3} \mbox{Td}(K3) \wedge ch(E_{\bf S}) = 2 \mbox{rk}(E_{\bf S}) + ch_2(E_{\bf S}) \ ,
\end{equation}
where
$\mbox{Td}(K3)$ is the Todd-class of $K3$, rk$(E)$ is the rank of the vector
bundle and  $ch_2(E_{\bf S})= -\tfrac{1}{2} \int \mbox{tr}_{\bf S}
{\mathcal F} \wedge {\mathcal F}$ is the second Chern-character.
Using $\mbox{rk}(\mbox{End} \ {\mathcal T}_{K3}) = 3$ we get
\begin{equation}\label{zero2}
\begin{aligned}
 h^{0,1} (\mbox{End} \ {\mathcal T}_{K3}) &= -6 + \tfrac{1}{2} \int
 \mbox{tr}_{\bf 3} ({\mathcal F} \wedge {\mathcal F}) 
= -6 + \tfrac{4}{2} \int \mbox{tr}_{\bf 2} ({\mathcal F} \wedge {\mathcal F}) 
= -6 + 4 \cdot 24 
= 90 \ ,
\end{aligned}
\end{equation}
where in the last step we used the integrated tadpole condition \eqref{BI2}
\begin{equation}
\tfrac12\int \mbox{tr}_{\bf 2} ({\mathcal F} \wedge {\mathcal F}) = \chi(K3) = 24\ .
\end{equation}

Summarizing, the Kaluza-Klein expansion of the gauge potential reads
\begin{equation}
 \dA_1 = V^{\bf 133} \ , \qquad \dA_{\bar 1} = C^{\bf 56} _j \omega_j
 \oplus \overline{C}^{\overline{\bf 56}} _j \overline{\omega}_j + \xi_k \alpha_k
 + \overline{\xi}_k \overline\alpha_k \ ,
\end{equation}
with $j=1,...,20$ and $k=1,...,90$.

\subsection{Coupling functions in the standard embedding}\label{potSE}
In this section we derive the coupling functions of the effective action. 
First we consider the kinetic terms in \eqref{Kinetic6Dterms} and in particular the couplings of the charged scalars. 
Due to the correspondence of their zero-modes to harmonic
$(1,1)$-forms \eqref{zero1} these functions exhibit a characteristic
dependence on the $K3$ moduli.\footnote{Recall that on $K3$ the embedding
$H^{1,1}(K3,\mathbb{R}) \subset H^2(K3, \mathbb{R})$ is a moduli dependent subspace.} 
To express this dependence in the following, let us review the parametrization of the $K3$ moduli space \eqref{K3mod} from \cite{Louis:2009dq}.
A Riemannian metric is given by a positive definite three-dimensional subspace $\Sigma := H^2 _+ (K3, \mathbb{R}) \subset H^2(K3,\mathbb{R})$, which is spanned by an orthonormal dreibein $(J_1, J_2, J_3)$. 
The $K3$ moduli $t^I _s$ are defined by the expansion
\begin{equation}\label{param}
 J_s = t^I _s \eta_I \ , \qquad I = 1,\ldots, 22 \ .
\end{equation}
They are constrained to be (positive) orthonormal
\begin{equation}
\qquad \rho_{IJ} t^I _s t^J _t = \delta_{st} \ , 
\end{equation}
and subject to an equivalence relation which identifies equivalent metrics
\begin{equation}
t^I _s \sim \tilde{t}^I _s = R_s ^{\ t} t^I _t \ , \qquad R \in SO(3) \ .
\end{equation}
$R$ rotates the dreibein inside $\Sigma$ and corresponds to an $S^2$ of possible complex structures per metric.

In the following we want to relate the moduli space of the charged scalars to the moduli space of $K3$ metrics.
Due to the very definition of ${\mathcal T}_{K3}$ in the standard embedding, the charged scalar zero modes are defined with respect to a chosen complex structure. 
Hence, the discussion of their couplings implicitly requires the breaking of the Hyperk\"ahler structure of $K3$.
Defining the complex structure via the 2-form $\Omega = J_1 + iJ_2$, 
the harmonic $(1,1)$ forms in the charged scalars zero modes \eqref{zero1} are given the projection
\begin{equation}\label{projector}
\eta_I ^{1,1} = (P^{1,1})_I ^{\ J} \eta_J \ , \qquad (P^{1,1})_I ^{\ J} = \delta_I ^J - \sum \limits_{s=1,2} \rho_{IK} t_s ^K t_s ^J \ ,
\end{equation}
where $\rho_{IJ}$ is the intersection form \eqref{insect}.
They depend on the complex structure moduli $t_1 ^I, t_2 ^I$.
In the following we fix the complex structure and discuss the dependence of the charged scalar couplings on the remaining K\"ahler moduli.
As in \eqref{zero1} $\eta_j, j=3,\ldots 22$ denotes a basis of $H^{1,1}(K3, \mathbb{R})$ with respect to the fixed complex structure.

%
%

Let us illustrate this by a first example. 
The KK reduction of \eqref{kin1} yields 
the kinetic term of the charged scalars 
in \eqref{Kinetic6Dterms}
\begin{equation}
-\tfrac{\alpha'}{\sqrt 2} {\mathcal D} \overline{C}^x _i \wedge * {\mathcal D}C^x _j \ {\mathcal V}^{-\frac32} \int g_{\bar{\alpha} \beta}
\overline{\omega}_{i} ^{\bar \alpha} \wedge \star \omega_j ^{\beta} \ ,
\end{equation}
where $g_{\bar{\alpha} \beta}$ is the K\"ahler metric on $K3$.
We show now that the charged scalar metric $G_{ij}$ is indeed related to the $b$-scalar metric $g_{IJ}$ given in \eqref{smetric}.
Using  the zero mode isomorphism \eqref{zero1} and the identities
\begin{equation}
 \Omega^{\bar{\alpha} \bar{\beta}} = f(z) |g|^{-\frac{1}{2}} \varepsilon^{\bar{\alpha} \bar{\beta}} \ , \qquad |f|^2 = \lVert \Omega \rVert^2  \sqrt{g} 
\end{equation}
as well as the normalization $\lVert \Omega \rVert^2 =
\tfrac2{\mathcal V}$
we obtain
\begin{equation}\label{metric1}
\begin{aligned}
 G_{ij} &= \tfrac{1}{\sqrt{2}{\mathcal V}^{\frac32}} \int g_{\bar{\alpha} \beta} \overline{\omega}_{i} ^{\bar \alpha} \wedge \star \omega_j ^{\beta} \\
&= \tfrac{\gamma_i \gamma_j}{\sqrt{2}{\mathcal V}^{\frac32} \lVert \Omega \rVert^{4}} \int g_{\bar{\alpha} \beta} g^{\delta \bar{\delta}} |f|^2 \varepsilon^{\bar{\alpha} \bar{\gamma}} \varepsilon^{\beta \gamma} (\eta_i)_{\bar{\gamma} \delta} 
(\eta_j)_{\gamma \bar{\delta}}  \lvert g \rvert^{-\tfrac12} d^4 x \\
&= \tfrac{\gamma_i \gamma_j}{\sqrt{2}{\mathcal V}^{\frac32} \lVert \Omega \rVert^2} \int g^{\bar{\gamma} \gamma} g^{\delta \bar{\delta}} (\eta_i)_{\bar{\gamma} \delta} 
(\eta_j)_{\gamma \bar{\delta}}  \sqrt{g} \ d^4 x \\
&= \tfrac{\gamma_i \gamma_j}{2 \sqrt{2 \mathcal V}} \int \eta_i \wedge \star \eta_j \ .
\end{aligned}
\end{equation}
From the last line in \eqref{metric1} (no summation over $i,j$ implied) one recognizes that this function is proportional to the projection of the $b$-scalar metric $g_{IJ}$
\begin{equation}\label{g}
g_{ij} := \int \eta_i \wedge \star \eta_j = (P^{1,1})_i ^{\ I} (P^{1,1})_j ^{\ J} g_{IJ} \ , \qquad g_{IJ} = \int \eta_I \wedge \star \eta_J  \ .
\end{equation}
While $P^{1,1}$ depends on the fixed complex structure, 
$g_{ij}$ also depends on the remaining K\"ahler moduli via the action of the Hodge $\star$ operator on $H^{1,1}(K3, \mathbb{R})$ \cite{Louis:2009dq}
\begin{equation}\label{star}
\star \eta_i = \Bigl(-\delta_i ^j + 2\rho_{ik} t^k _3 t^j _3 \Bigr) \eta_j \ .
\end{equation}

%

For the coupling function $N^I _{ij}$ in \eqref{N} 
which is obtained from a KK reduction of \eqref{HH}
we first use the same manipulations as above to get
\begin{equation}\label{N3}
\begin{aligned}
N_{ij}\ =\ \Omega_{\alpha \beta} \ \omega^{\alpha} _i \wedge \omega^{\beta} _j\
 &=\ \tfrac{\gamma_i \gamma_j}{\lVert \Omega \rVert^2} \overline{\Omega} \cdot (\eta_i \wedge \eta_j) \
 =\ \tfrac{\gamma_i \gamma_j}{\lVert \Omega \rVert^2} \rho_{ij}
 \overline\Omega\cdot \mbox{vol}\ 
 =\  \tfrac{\mathcal V \gamma_i \gamma_j}{2 \sqrt{g}} \rho_{ij} \overline{\Omega} \ .
\end{aligned}
\end{equation}
Here $\cdot$ denotes the contraction of forms and $\mbox{vol}$ is the volume form, normalized to 1.
In the second step we used  $\eta_i \wedge \eta_j= \rho_{ij} \mbox{vol}$ and in the third step we used 
$\overline{\Omega}\cdot\mbox{vol} = g^{-\frac12} \overline{\Omega}$.
The coupling  $\rho_{ij}$ is defined as the projection 
\begin{equation}
\begin{aligned}
\rho_{ij}&:= \int \eta_i \wedge  \eta_j = (P^{1,1})_i ^{\ I} (P^{1,1})_j ^{\ J} \rho_{IJ} \ ,
\end{aligned}
\end{equation}
where $\rho_{IJ}$ is the moduli independent intersection matrix.
Hence, the expansion into $\eta_I$ has coefficients 
\begin{equation}\label{NIijeq} 
\begin{aligned}
 N^I _{ij} &= \rho^{IJ} \int N_{ij} \wedge \eta_J 
= \tfrac12 \gamma_i \gamma_j \rho_{ij} \rho^{IJ} \int \overline{\Omega} \wedge \eta_J 
= \tfrac12 \gamma_i \gamma_j \rho_{ij} \rho^{IJ} (\langle J_1, \eta_J \rangle -i \langle J_2, \eta_J\rangle ) \ ,
\end{aligned}
\end{equation}
where $\langle \cdot, \cdot \rangle$ is the scalar product 
on $H^2(K3, \mathbb{R})$.

For the coupling function $M^I _{ij}$ in \eqref{N} which also arise
from
\eqref{HH} we proceed
similarly to get
\begin{equation}\label{N1}
 M_{ij} = \tfrac{1}{\sqrt{2 \mathcal V}} g_{\alpha \bar{\beta}} \omega_i ^{\alpha} \wedge \overline{\omega}_j ^{\bar \beta} = 
\tfrac{\sqrt{\mathcal V} \gamma_i \gamma_j}{2\sqrt{2}} g^{\gamma \bar{\delta}}
(\eta_i)_{\gamma \bar{\alpha}} (\eta_j)_{\beta \bar{\delta}} dz^{\beta} \wedge d{\overline z}^{\bar \alpha} \ .
\end{equation}
Identifying the components of the K\"ahler form as $g_{\alpha \bar{\beta}} = -iJ_{\alpha \bar{\beta}}$ and $g_{\bar{\alpha} \beta} = iJ_{\bar{\alpha} \beta}$ we can express $M_{ij}$ as the 
special contraction
\begin{equation}\label{N4}
\begin{aligned}
 M_{ij} &= -i \tfrac{\sqrt{\mathcal V} \gamma_i \gamma_j}{2\sqrt{2}} \Bigl( J \cdot (\eta_i \wedge
 \eta_j) - (J \cdot \eta_i) \ \eta_j - ( J \cdot \eta_j) \ \eta_i \Bigr) \\
&= -i \tfrac{\sqrt{\mathcal V} \gamma_i \gamma_j}{2\sqrt{2}} \Bigl( \rho_{ij} (J \cdot \mbox{vol}) - \tfrac{\sqrt{2 \mathcal V}}{\sqrt g} \langle J_3, \eta_i \rangle \eta_j - \tfrac{\sqrt{2 \mathcal V}}{\sqrt g} 
\langle J_3, \eta_j \rangle \eta_i \Bigr) \\
&= -i \tfrac{\mathcal V \gamma_i \gamma_j}{2\sqrt g} \Bigl( \rho_{ij} J_3 - \langle J_3, \eta_i \rangle \eta_j - \langle J_3, \eta_j \rangle \eta_i \Bigr) \ .
\end{aligned}
\end{equation}
Here we used the following identities
\begin{equation}
 \begin{aligned}
  (J \cdot \eta_i) \mbox{vol} &= \tfrac{1}{\sqrt g} J \wedge \eta_i = \sqrt{\tfrac{2 \mathcal V}{g}} \langle J_3, \eta_i \rangle \mbox{vol} \ , \qquad
 J \cdot \mbox{vol} = \tfrac{1}{\sqrt g} J \ .
 \end{aligned}
\end{equation}
 Hence, the expansion into $\eta_I$ has coefficients 
\begin{equation}\label{N2}
 M^I _{ij} = \rho^{IJ} \int M_{ij} \wedge \eta_J
 = i\tfrac{\gamma_i \gamma_j}{2} \rho^{IJ} \Bigl( \rho_{ij} \langle J_3, \eta_J \rangle - \langle J_3, \eta_i \rangle \rho_{jJ} - \langle J_3, \eta_j \rangle \rho_{iJ} \Bigr) \ .
\end{equation}

Both couplings $M$ and $N$ depend on the $K3$ moduli but for a fixed complex structure we have the following simplification. 
In a basis $(\eta_1, \eta_2, \eta_i)$ of $H^2(K3, \mathbb{R})$, where $\eta_{1,2}$ span the complex structure 2-plane, 
we have $\langle J_{1,2}, \eta_I \rangle = 0$ for $I=i$ and $\langle J_3, \eta_I \rangle = 0$ for $I=1,2$.
This implies
\begin{equation}
\begin{aligned}
 N_{ij} ^I \neq 0 \quad &\mbox{only for} \ I=1,2 \ , \\
 M_{ij} ^I \neq 0 \quad &\mbox{only for} \ I=3,\ldots,22 \ . 
\end{aligned}
\end{equation}
In this basis the couplings \eqref{Db} between the charged scalars and the $b$-scalars reduce to
\begin{equation}\label{Db2}
 {\mathcal D}_c b^I = \begin{pmatrix} db^{1,2} - \alpha' \varepsilon_{xy} (N^{1,2} _{kl} C^x _k {\mathcal D} C^y _l + c.c.) - \ldots \\
   db^i - \alpha' \delta_{xy} M^i _{kl} \overline{C}^x _{k} \overleftrightarrow{\mathcal D}  C^y _l - \ldots \end{pmatrix} \ ,
\end{equation}
where the dots stand for the $\overline{\xi} d \xi$ terms.
Moreover, for the $b$-scalar combination $b^i \eta_i = t_3 ^i \eta_i$ proportional to the K\"ahler form of $K3$, the coupling function reduces to
\begin{equation}
 M_{ij} = -i \tfrac{\gamma_i \gamma_j}{2} g_{ij} \ ,
\end{equation}
with $g_{ij}$ known from \eqref{g}.
In appendix \ref{limit} we will use the second row in \eqref{Db2} to identify a quaternionic K\"ahler moduli subspace, containing complexified K\"ahler moduli and charged scalars.

Let us now turn to the scalar potential which contains quartic terms
of the charged scalars. These arise from the squares of the expressions 
\eqref{problem} and \eqref{Dse}. The term in \eqref{Dse}, which is in
the adjoint representation of the surviving gauge group, gives rise to
$D$-terms in 6D. The term in \eqref{problem}
is not allowed by 6D supergravity
and we shall prove here that it vanishes due to properties of $K3$ and its bundles. First recall from \eqref{Vym} that only the selfdual components $\delta F_{{\bar 2}+}$ contribute 
to the scalar potential which will be crucial to show the  
consistency with 6D supergravity. Recall  \eqref{problem}
\begin{equation}\label{problem1}
 \Fl_{\bar 2} ^{({\bf 1},{\bf 3})} = \begin{pmatrix} \overline{C}^x _i \\ C^x _i \end{pmatrix}^T \   
\begin{pmatrix} \sigma^s _{\bar{\alpha} \beta} \overline{\omega} ^{\bar \alpha} _i \wedge \omega^{\beta} _j & \sigma^s _{\bar{\alpha} \bar{\beta}} \ \overline{\omega}^{\bar \alpha} _i \wedge \overline{\omega}^{\bar \beta} _j 
\\ \sigma^s _{\alpha \beta} \ \omega^{\alpha} _i \wedge \omega^{\beta} _j & \sigma^s _{\alpha \bar{\beta}} \ \omega^{\alpha} _i \wedge \overline{\omega}^{\bar \beta} _j \end{pmatrix} \delta_{xy} \begin{pmatrix} C^y _j \\ \overline{C}^y _j \end{pmatrix} \ ,
\end{equation}
where all matrix elements are 2-forms in the group 
$H^2(\mbox{End} \ {\mathcal T}_{K3})$ as follows from the group
representation ${({\bf 1},{\bf 3})}$.
We now use a local decomposition of $H^2(\mbox{End} \ {\mathcal
  T}_{K3})$ and show that its 
global extension does not exist.
In fact any 2-form in $H^2(\mbox{End} \ {\mathcal T}_{K3})$ 
can be locally trivialized as
\begin{equation}
 f^i \otimes \omega_i \ \in \ \Gamma(\mbox{End} \ {\mathcal T}_{K3}) \otimes \Lambda^2(K3) \ , 
\end{equation}
where $i=1,...,6$. Since 
the zero modes in \eqref{problem1} are  $d_{\mathcal A}$-closed
also their products 
are $d_{\mathcal A}$-closed.
This implies
\begin{equation}\label{arg}
 0 = d_{\mathcal A} (f^i \otimes \omega_i) = (d_{\mathcal A} f^i) \wedge \omega_i + f^i (d\omega_i) \ .
\end{equation}
For the scalar potential we restrict this equation to the selfdual 2-forms. Since there exists on $K3$ a basis of $d$-closed selfdual 2-forms,
\eqref{arg} reduces in this basis to
\begin{equation}\label{sol}
 d_{\mathcal A} f^i = 0 \ .
\end{equation}
Hence, the $f^i$ are covariantly constant sections of $\mbox{End} \
{\mathcal T}_{K3}$, which have to extend to globally constant
sections. However,
since $\mbox{End} \
{\mathcal T}_{K3}$ is an irreducible, 
nontrivial bundle, only the constant zero section exists. In other words, the deformation \eqref{problem1} preserves the ASD property of the background field strength and therefore 
does not contribute to the scalar potential.

Next we calculate the selfdual part of \eqref{Dse}
\begin{equation}\label{Dse2}
\Fl_{\bar 2} ^{({\bf 133},{\bf 1})} = \begin{pmatrix} \overline{C}^x _i \\ C^x _i \end{pmatrix}^T \   \begin{pmatrix} \tfrac{1}{\sqrt{2\mathcal V}} g_{\bar{\alpha} \beta} \ 
\overline{\omega} ^{\bar \alpha} _i \wedge \omega^{\beta} _j & \overline{\Omega}_{\bar{\alpha} \bar{\beta}} \ \overline{\omega}^{\bar \alpha} _i \wedge 
\overline{\omega}^{\bar \beta} _j \\ \Omega_{\alpha \beta} \ \omega^{\alpha} _i \wedge \omega^{\beta} _j & \tfrac{1}{\sqrt{2\mathcal V}} g_{\alpha \bar{\beta}} \ \omega^{\alpha} _i 
\wedge \overline{\omega}^{\bar \beta} _j \end{pmatrix} (\tau^a) _{xy} \begin{pmatrix} C^y _j \\ \overline{C}^y _j \end{pmatrix} \ . 
\end{equation}
We recognize that the same coupling functions appear as 
in \eqref{N3} and \eqref{N1} so that we have
\begin{equation}
 \Fl_{\bar 2} ^{({\bf 133},{\bf 1})} =  \begin{pmatrix} \overline{C}^x _i \\ C^x _i \end{pmatrix}^T \ \begin{pmatrix} -M_{ij} & \overline{N}_{ij} \\ N_{ij} & 
M_{ij} \end{pmatrix} (\tau^a) _{xy} \begin{pmatrix} C^y _j \\ \overline{C}^y _j \end{pmatrix} \ .
\end{equation}
The off-diagonal elements are already selfdual 2-forms given by \eqref{N3}, while the diagonal elements are generic $(1,1)$-forms. 
We get their selfdual part by projecting onto $J_3$
\begin{equation}
\begin{aligned}
 M_{ij+} 
&= \left(\int M_{ij} \wedge J_3 \right) J_3 
= -i\tfrac{\gamma_i \gamma_j}{2} \Bigl( \rho_{ij} \langle J_3, J_3 \rangle -2 \langle J_3, \eta_i \rangle \langle J_3, \eta_j \rangle \Bigr) J_3 
= i\sqrt{2 \mathcal V} G_{ij} J_3 \ .
\end{aligned}
\end{equation}
Here we identified the kinetic coupling $G_{ij}$ using \eqref{star}, \eqref{metric1} and
\begin{equation}
 g_{ij} = \int \eta_i \wedge \star \eta_j = (-\delta_j ^k + 2\rho_{jl}t_3 ^k t_3 ^l) \rho_{ik} 
= -\rho_{ij} \langle J_3, J_3 \rangle +2 \langle J_3, \eta_i \rangle \langle J_3, \eta_j \rangle \ .
\end{equation}
Summarizing, we have 
\begin{equation}\label{Dse++}
 \Fl_{\bar{2}+} ^{({\bf 133},{\bf 1})} = \begin{pmatrix} \overline{C}^x _i \\ C^x _i \end{pmatrix}^T  \begin{pmatrix} -i\sqrt{2 \mathcal V} G_{ij} J_3 & \tfrac12 \tilde{\rho}_{ij} \Omega 
\\ \tfrac12 \tilde{\rho}_{ij} \overline{\Omega} & i\sqrt{2 \mathcal V} G_{ij} J_3 \end{pmatrix} 
(\tau^a)_{xy} \begin{pmatrix} C^y _j \\ \overline{C}^y _j \end{pmatrix} \ ,
\end{equation}
where we $\tilde{\rho}_{ij} = \gamma_i \gamma_j \rho_{ij}$ denotes the rescaled intersection matrix on $H^{1,1}(K3, \mathbb{R})$.

\subsection{Zero modes in line bundle backgrounds}\label{line}
We now apply the results from appendix~\ref{def1} to deformations of a line bundle background. For one $U(1)$ principal bundle inside one $E_8$ factor we have the breaking 
\begin{equation}
 E_8 \longrightarrow G \times \langle U(1) \rangle \ ,
\end{equation}
and the adjoint decomposition
\begin{equation}
 {\bf 248} \longrightarrow \bigoplus \limits_i \ \left(({\bf R}_i,{\bf 1}_{q_i}) \oplus ({\overline{\bf R}}_i,{\bf 1}_{-q_i})\right) \oplus ({\mathfrak g}, {\bf 1}_0) \oplus ({\bf 1}, {\bf 1}_0) \ ,
\end{equation}
which defines the associated vector bundles. Due to \eqref{vec} we get
again one 6D gauge potential $\V^{\mathfrak g}$ in the adjoint of
$G$. However, now 
the $\langle U(1) \rangle$ is part 
of the unbroken gauge group since it commutes with itself. Since here ${\mathfrak h} = {\bf 1}_0$ corresponds to the trivial line bundle, there also 
exists a 6D Abelian gauge potential $\V^{\bf 1}$ in the same representation $({\bf 1}, {\bf 1}_0)$ as the background connection $\mathcal A$.
There exist no bundle moduli, since $\mbox{End} \ L^q = {\mathcal O}$ 
is the trivial bundle and
\begin{equation}
 H^{0,1} (\mbox{End} \ L^q) \cong H^{0,1} (K3, \mathbb{R}) = 0 \ .
\end{equation}
Finally, we get charged scalars in representations ${\bf R}_i$. Their multiplicity cannot be related to the Hodge numbers of $K3$, but we have
\begin{equation}\label{chi3}
 h^{0,1} (L^q) = -\chi(L^q) \ ,
\end{equation}
by the same argument as in \eqref{index}.
The chiral index of a line bundle over a four-dimensional manifold
takes the simplified form \eqref{chiline} as we will show now.
The total Chern-character 
$ch(L)={\rm tr}\exp(\frac{i}{2\pi}{\mathcal F})$ factorizes for product bundles,
\begin{equation}
 ch(L^q) = ch(L) \wedge ... \wedge ch(L) \ , 
\end{equation}
which implies
\begin{equation}
ch_2(L^q) = q \ ch_2(L) + \tfrac{1}{2} q(q-1) ch_1(L)^2 \ .
\end{equation}
For line bundles we have $ch_2(L) = \tfrac{1}{2} ch_1(L)^2$ such that
\begin{equation}
 ch_2(L^q) = q^2 ch_2(L) \ .
\end{equation}
Using $\mbox{rk}(L^q) = \mbox{rk}(L) = 1$, the chiral index reduces to
\begin{equation}
 \chi(L^q) = 2\mbox{rk}(L^q) + ch_2(L^q) = 2 + q^2 ch_2(L) \ .
\end{equation}
Therefore \eqref{chiline} is verified.

The Kaluza-Klein expansion of the gauge potential is analogous to
\eqref{agenau} and reads
\begin{equation}
\dA_1 = \V^{\mathfrak g} + \V^{\bf 1} \ , \qquad 
\dA_{\bar 1} = \sum_i \ (C_{k_i} ^{{\bf R}_i} \omega_{k_i} ^{q_i} + \overline{C}_{k_i} ^{\overline{\bf R}_i} \overline{\omega}_{k_i} ^{-q_i})  + (\overline{D}_{k_i} ^{\overline{\bf R}_i} \overline{\varpi}_{k_i} ^{-q_i} + D_{k_i} ^{{\bf R}_i} \varpi_{k_i} ^{q_i}) \ .
\end{equation}
The zero modes belong to the Dolbeault cohomology groups
\begin{equation}
\begin{aligned}
 \omega_{k_i} ^{q_i} \in H^{0,1} (L^{q_i}) \ &, \ \ \ \ 
 \overline{\omega}_{k_i} ^{-q_i} \in H^{1,0} (L^{-q_i}) \ , \\
 {\varpi}_{k_i} ^{q_i} \in H^{1,0} (L^{q_i}) \ &, \ \ \ \ 
 \overline{\varpi}_{k_i} ^{-q_i} \in H^{0,1} (L^{-q_i}) \ ,
\end{aligned}
\end{equation}
with multiplicities $k_i = 1,..., -\chi(L^{q_i})$.

The scalar potential of the charged scalars contains the selfdual
parts of \eqref{Dline1}, \eqref{Dline2} and \eqref{problem2}, i.e.
\begin{equation}
 \Fl_{\bar{2}+} ^{\mathfrak g} \ , \quad \Fl_{\bar{2}+} ^{\bf 1} \ , \quad \Fl_{\bar{2}+} ^{{\bf R}_i \oplus \overline{\bf R}_i} \ .
\end{equation}
We show first that any term of the form $\Fl_{\bar{2}+} ^{{\bf R}_i \oplus \overline{\bf R}_i}$ vanishes. 
The product of internal zero modes in \eqref{problem2} belong to $H^2 (L^{q_i} \oplus L^{-q_i})$
and they are also closed under the gauge covariant derivative $d_{\mathcal A}$. Locally we can write these 2-forms as 
\begin{equation}
 s^i \otimes \alpha_i \ , \ \ s^i \in \Gamma(L^{q_i} \oplus L^{-q_i}) \ , \ \ \alpha_i \in \Lambda^2(K3) \ ,
\end{equation}
where $i = 1,...,6$ is the number of locally independent 2-forms. Then we have
\begin{equation}\label{2form}
 0 = d_{\mathcal A}(s^i \otimes \alpha_i) = (d_{\mathcal A} s^i) \wedge \alpha_i + s^i \otimes (d\alpha_i) \ .
\end{equation}
If we restrict to the $d$-closed selfdual 2-forms, \eqref{2form} reduces to 
\begin{equation}
 0 = (d_{\mathcal A} s^j) \wedge \alpha_j ^+ \ .
\end{equation}
It follows that $\Fl_{\bar{2}+} ^{{\bf R}_i \oplus \overline{\bf R}_i}$ is proportional to covariantly constant sections 
$s^j \in \Gamma(L^{q_i} \oplus L^{-q_i})$. However, since $L^{q_i} \oplus L^{-q_i}$ is nontrivial and irreducible, only the constant zero section exists. 
We conclude that all $\Fl_{\bar{2}+} ^{{\bf R}_i \oplus \overline{\bf R}_i}$ vanish.

Next we derive the selfdual part of $\Fl_{\bar{2}+} ^{\mathfrak g}$ and $\Fl_{\bar{2}+} ^{\bf 1}$. 
Considering the matrix of internal 2-forms in \eqref{Dline1} and \eqref{Dline2},
\begin{equation}
 \begin{pmatrix} \overline{\omega}^{-q_i} _{k_i} \wedge \omega^{q_i} _{l_i} & \overline{\omega}^{-q_i} _{k_i} \wedge \varpi^{q_i} _{l_i} \\ \overline{\varpi}^{-q_i} _{k_i} \wedge \omega^{q_i} _{l_i} & \overline{\varpi}^{-q_i} _{k_i} \wedge \varpi^{q_i} _{l_i} \end{pmatrix} \ ,
\end{equation}
they take values in the trivial bundle, $H^2(K3,L^{q_i} \otimes L^{-q_i}) = H^2(K3)$. Hence, covariantly constant sections exist.
Projecting to the selfdual components we get
\begin{equation}
\begin{aligned}
 (\overline{\omega}^{-q_i} _{k_i} \wedge \omega^{q_i} _{l_i})_+ &= \tfrac{i}{2 \mathcal V} \left( \int \overline{\omega}^{-q_i} _{k_i} \wedge \star \omega^{q_i} _{l_i} \right) J \ , \\
 (\overline{\varpi}^{-q_i} _{k_i} \wedge \varpi^{q_i} _{l_i})_+ &= \tfrac{i}{2 \mathcal V} \left( \int \overline{\varpi}^{-q_i} _{k_i} \wedge \star \varpi^{q_i} _{l_i} \right) J \ , \\
 (\overline{\omega}^{-q_i} _{k_i} \wedge \varpi^{q_i} _{l_i})_+ &= \tfrac{1}{2} \left( \int (\overline{\omega}^{-q_i} _{k_i} \wedge \varpi^{q_i} _{l_i}) \wedge \overline{\Omega} \right) \Omega \ , \\
 (\overline{\varpi}^{-q_i} _{k_i} \wedge \omega^{q_i} _{l_i})_+ &= \tfrac{1}{2} \left( \int (\overline{\varpi}^{-q_i} _{k_i} \wedge \omega^{q_i} _{l_i}) \wedge {\Omega} \right) \overline{\Omega} \  .
\end{aligned}
\end{equation}
The diagonal elements are proportional to the scalar kinetic metric $g_{k_i l_i} ^C$ and $g_{k_i l_i} ^D$, that appeared in \eqref{skinetic}.
The off-diagonal elements contain a generalized intersection matrix
\begin{equation}
 c_{k_i l_i} = \int (\overline{\omega}^{-q_i} _{k_i} \wedge \varpi^{q_i} _{l_i}) \wedge \overline{\Omega} \ ,
\end{equation}
where the indices run over the multiplicity of the corresponding charged scalars.

\section{$T^4/\mathbb{Z}_3$ limit: Hypermultiplet moduli space metric}\label{limit}

In this appendix we focus on a specific orbifold corresponding to a 
heterotic compactification on a smooth $K3$ with standard embedding
for the gauge bundle. In this case we are able to give an explicit form of the
hypermultiplet field space for the untwisted moduli.

 Specifically we consider the $E_8\times E_8$
heterotic 
string  compactified on the orbifold $T^4/\mathbb{Z}_3$ with gauge
twist given by $\frac13 (1^2,0^6)(0^8)$ \cite{Erler:1993zy}.
In this case the unbroken gauge group is $E_7\times U(1)\times E_8$. In the hypermultiplet sectors we have both untwisted and twisted states in the following representations:\footnote{The untwisted spectrum is obtained by taking the spectrum coming from compactification on $T^4$ and performing the $\mathbb{Z}_3$ projection. The twisted spectrum comes from strings localized around the orbifold singularities.}
\begin{equation}
 ({\bf 56,1})_1^{\rm untw} \oplus ({\bf 1,1})_2^{\rm untw} \oplus 2 ({\bf 1,1})_0^{\rm untw} \oplus  9({\bf 56,1})_{\frac13}^{\rm tw} \oplus 45 ({\bf 1,1})_{\frac23}^{\rm tw} \oplus 18 ({\bf 1,1})_{\frac43}^{\rm tw} \:.
\end{equation}

When we blow up the orbifold $T^4/\mathbb{Z}_3$ we get a smooth
$K3$. After a field redefinition, the orbifold spectrum matches with
the spectrum obtained by a smooth compactification with nontrivial
gauge bundle \cite{Honecker:2006qz}. In particular, the two $({\bf
  1,1})_0^{\rm untw}$ are the two hypermultiplets containing the four
geometric moduli and the four $B$-field moduli surviving the
$\mathbb{Z}_3$ projection, the $({\bf 56,1})_1^{\rm untw}$ is a
charged field, and the $({\bf 1,1})_2^{\rm untw}$ is eaten to give
mass to the $U(1)$ gauge boson. 
Therefore the total orbifold spectrum matches the spectrum of the
smooth compactification considered in section~\ref{SEE}, i.e.~20 geometric, 45 bundle moduli and 10 charged hypermultiplets.

The metric on the hypermultiplet scalar field space in the untwisted sector, can be obtained by considering the 6D heterotic compactification on $T^4$ and performing a 
suitable truncation \cite{Witten:1985xb,Ferrara:1986qn}. For the case at hand the truncation is
\begin{equation}
 \frac{SO(4,4+N)}{SO(4)\times SO(4+N)} \qquad \rightarrow \qquad
  \frac{SU(2,2+n)}{U(1)\times SU(2)\times SU(2+n)}\ .
\end{equation}
The latter space is simultaneously quaternionic-K\"ahler and K\"ahler,
with a metric determined by the K\"ahler potential
\begin{equation}\label{Kpot}
K=-\log \det (T+T^\dagger - 2\Psi \Psi^\dagger) \:.
\end{equation}
$\Psi$ is a $2\times n$ complex matrix, which encodes the two complex
scalars belonging to the $n$ hypermultiplets in the untwisted charged
spectrum
(in our case $n=56$.) $T$ is a $2\times 2$ complex matrix given by
\begin{equation}\label{TijApp}
(T_{ij}) = \left(\begin{array}{ccc}
 g_{1\bar{1}}+ i B_{1\bar{1}} + \Psi_1 \overline{\Psi}_1 && g_{12}+ i B_{12} + \Psi_1 \overline{\Psi}_2\\
 \overline{g_{12}} + i \overline{B_{12}} + \Psi_2 \overline{\Psi}_1 && g_{2\bar{2}}+ i B_{2\bar{2}} + \Psi_2 \overline{\Psi}_2\\
\end{array}\right) \:.
\end{equation}
It contains the real $g_{1\bar{1}}$, $g_{2\bar{2}}$ and the complex $g_{12}$ metric elements and the the corresponding components of the $B$-field. $\Psi_i \overline{\Psi}_j$ includes a summation over the $n$ components.
For simplicity let us fix the complex structure such that $g_{12}=0$. In this limit, the K\"ahler potential \eqref{Kpot} yields the kinetic terms
\begin{equation}
 K_{T_{ij} \overline{T}_{kl}} dT_{ij} d{\overline T}_{kl} = \tfrac{1}{4g_{1 \bar{1}} ^2} dT_{11} d{\overline T}_{11} + \tfrac{1}{4g_{2 \bar{2}} ^2} dT_{22} d{\overline T}_{22} + 
\tfrac{1}{4g_{1 \bar{1}}g_{2 \bar{2}}} (dT_{12} d{\overline T}_{12} + dT_{21} d{\overline T}_{21}) \ , 
\end{equation}
\begin{equation}
 K_{\Psi_i {\overline \Psi}_j} d\Psi_i d{\overline \Psi}_j = (\tfrac{1}{g_{1 \bar{1}}} + \tfrac{\Psi_2 \overline{\Psi}_2}{g_{1 \bar{1}} g_{2 \bar{2}}} + 
\tfrac{\Psi_1 \overline{\Psi}_1}{g_{1 \bar{1}} ^2}) d\Psi_1 d{\overline \Psi}_1 + (\tfrac{1}{g_{2 \bar{2}}} + \tfrac{\Psi_1 \overline{\Psi}_1}{g_{1 \bar{1}} g_{2 \bar{2}}} + 
\tfrac{\Psi_2 \overline{\Psi}_2}{g_{2 \bar{2}} ^2}) d\Psi_2 d{\overline \Psi}_2 \ ,
\end{equation}
\begin{equation}
 K_{T_{ij} \overline{\Psi}_k} dT_{ij} d{\overline \Psi}_k = -\tfrac{\Psi_1}{2g_{1 \bar{1}} ^2} dT_{11} d{\overline \Psi}_1 - \tfrac{\Psi_2}{2g_{2 \bar{2}} ^2} dT_{22} d{\overline \Psi}_2
-\tfrac{\Psi_2}{2g_{1 \bar{1}} g_{2 \bar{2}}} dT_{12} d{\overline \Psi}_1 - \tfrac{\Psi_1}{2g_{1 \bar{1}} g_{2 \bar{2}}} dT_{21} d{\overline \Psi}_2 \ .
\end{equation}
Inserting \eqref{TijApp} we get the kinetic terms in terms of the Kaluza-Klein modes \cite{Witten:1985xb,Ferrara:1986qn}. The leading term for the charged scalars reads
\begin{equation}\label{CC}
 \sum \limits_{i=1,2} \tfrac{1}{g_{i \bar i}} d\Psi_i d{\overline \Psi}_i \ .
\end{equation}
The terms for the two complexified K\"ahler moduli read
\begin{equation}\label{bb}
 \sum \limits_{i=1,2} \tfrac{1}{4g_{i \bar i} ^2} \lvert dg_{i \bar i} + idB_{i \bar i} + {\overline \Psi}_i d \Psi_i - \Psi_i d {\overline \Psi}_i \rvert^2 \ .
\end{equation}
The terms for the off-diagonal fields in $T$ read
\begin{equation}\label{bC}
 \tfrac{1}{4g_{1 \bar 1}g_{2 \bar 2}} \bigl( \lvert idB_{12} + \Psi_1 d{\overline \Psi}_2 - {\overline \Psi}_2 d\Psi_1 \rvert^2 + \lvert id\overline{B_{12}} + \Psi_2 d{\overline \Psi}_1 - {\overline \Psi}_1 d \Psi_2 \rvert^2 \bigr) \ .
\end{equation}

We now compare the above kinetic couplings with our results \eqref{Kinetic6Dterms} coming from the smooth $K3$.
%
%
To make contact with the ones just derived, we have to take the orbifold limit and identify the $K3$ moduli related to $g_{i\bar{i}}$.
The $T^4/\mathbb{Z}_3$ limit of $K3$ corresponds to taking the 3-plane $\Sigma$ orthogonal to 18 two-cycles with intersection matrix $A_2^{\oplus 9}$.\footnote{
$T^4/\mathbb{Z}_3$ has nine $A_2$-singularities (i.e. locally $\mathbb{C}^2/\mathbb{Z}_3$). One ADE singularity of $K3$ is generated by shrinking a set of two-cycles with the intersection matrix given by (minus) the Cartan matrix of the corresponding ADE group.
} 
The orthogonal complement (where $\Sigma$ lives) must contain the two complex 2-tori (that we call $\eta_1,\eta_2$) spanned by the coordinates $z^i$, plus two 2-cycles (called $\eta_3,\eta_4$) with positive self-intersection and that are not of type $(1,1)$. They have the following intersection matrix:
\begin{equation}
\left(\begin{array}{cccc}
  0 & 3 && \\ 3& 0 &&\\ && 2 & -1\\ && -1 & 2
\end{array}\right)\ .
\end{equation}

The chosen complex structure (i.e. $g_{12}=0$) makes the metric hermitean, allowing us to identify the $g_{i\bar{i}}$ elements with the coefficient of the K\"ahler form $J$ along the Poincar\'e dual of the two 2-tori.
On the K3 side we need to take the two 2-tori of type $(1,1)$. This is done by making  $J$ be a linear combination of (the Poincar\'e dual of) $\eta_1$ and $\eta_2$ and $\Omega$ live in the positive definite subspace $\{\eta_3,\eta_4\}$. Also $B$ will have components along $\eta_1$ and $\eta_2$:
\begin{equation}
 J = t^1 \, \eta_1 + t^2 \, \eta_2 \ ,\qquad\qquad   B = b^1 \, \eta_1
 + b^2 \, \eta_2 + ...\ ,
\end{equation}
and we have the identifications $g_{i\bar{i}} \leftrightarrow t^i$ and $B_{i\bar{i}} \leftrightarrow b^i$.

First, consider the coupling in front of \eqref{bb}. 
The smooth result reduces in the orbifold limit to
\begin{equation}
 \tfrac{1}{\mathcal V} g_{IJ} = \tfrac{1}{\mathcal V} \int \eta_I \wedge \star \eta_J \ \longrightarrow \begin{pmatrix} \frac{1}{(t^1)^2} & \\ & \frac{1}{(t^2)^2} \end{pmatrix} \ ,
\end{equation}
which matches with \eqref{bb} up to a numerical constant. For the leading charged scalar coupling we have
\begin{equation}
 G_{ij} = \tfrac{\gamma_i \gamma_j}{2\sqrt{2 \mathcal V}} \int \eta_i \wedge \star \eta_j \ \longrightarrow \ \tfrac{3}{2\sqrt{2}} 
\begin{pmatrix} \tfrac{1}{\langle J,\eta_1 \rangle} \frac{t^2}{t^1} &  \\  & \tfrac{1}{\langle J,\eta_2 \rangle} \frac{t^1}{t^2} \end{pmatrix} 
=  \tfrac{1}{2\sqrt{2}} \begin{pmatrix} \frac{1}{t^1} &  \\  & \frac{1}{t^2} \end{pmatrix} \ ,
\end{equation}
which matches with \eqref{CC}.
Here we see that for the orbifold match it is necessary to include the moduli dependent functions $\gamma_j = {\mathcal V}^{\frac{1}{4}}/\langle J, \eta_i \rangle^{\frac12}$ 
in the isomorphy of zero modes \eqref{zero1}. 
In fact, the moduli dependence of the skew-symmetric couplings $M_{ij} ^I$ drops out in the orbifold limit, as expected.
The only nonvanishing components are
\begin{equation}
 M_{11} ^1 = M_{22} ^2 = -\tfrac{i}{\sqrt 2} \ .
\end{equation}
This matches with \eqref{bC}.

\end{appendix}

\end{document}